\documentclass[journal,onecolumn,12pt]{IEEEtran}
\IEEEoverridecommandlockouts

\usepackage[utf8]{inputenc}
\usepackage[T1]{fontenc}
\usepackage[american]{babel}
\usepackage{graphicx}
\usepackage{cite}
\usepackage{listings}
\usepackage{float}
\usepackage{amsmath,amssymb,exscale}
\usepackage{lettrine}
\usepackage{subcaption}
\usepackage{blindtext, graphicx}
\usepackage{verbatim}
\usepackage{algorithm}
\usepackage{algpseudocode}
\usepackage{fancyvrb}
\usepackage{bera}
\usepackage{mathtools}
\usepackage{lipsum}

\usepackage{times}
\usepackage{scalerel}
\usepackage{tikz}
\usetikzlibrary{svg.path}
\usepackage[bookmarks=false]{hyperref}
\usepackage{caption}

\newcommand{\revise}{\textcolor{black}}

\begin{document}
   \author{
    \IEEEauthorblockN{Jun Wu, \IEEEmembership{Graduate Student Member, IEEE}}, \IEEEauthorblockN{Weijie Yuan, \IEEEmembership{Member, IEEE}}, \IEEEauthorblockN{Zhiqiang Wei, \IEEEmembership{Member, IEEE}}, \IEEEauthorblockN{Kecheng Zhang}, \IEEEauthorblockN{Fan Liu, \IEEEmembership{Senior Member, IEEE}}, and \IEEEauthorblockN{Derrick Wing Kwan Ng, \IEEEmembership{Fellow, IEEE}}
    \thanks{
    % This work was supported in part by the National Natural Science Foundation of China under Grants 62101232 and 62201242; in part by Guangdong Provincial Natural Science Foundation under Grant 2022A1515011257; in part by Shenzhen Science and Technology Program under Grant JCYJ20220530114412029. (Corresponding author: \textit{Weijie Yuan})
 Part of the paper was
presented at ICASSP 2024 \cite{wu2023optimal}.
(\textit{Corresponding author: Zhiqiang Wei})

J. Wu, W. Yuan, K. Zhang, and F. Liu are with the School of System Design and Intelligent Manufacturing, Southern University of Science and Technology, Shenzhen 518055, China (email: wuj2021@mail.sustech.edu.cn; 
 yuanwj@sustech.edu.cn; 
 zhangkc2022@mail.sustech.edu.cn; liuf6@sustech.edu.cn).

Z. Wei is with the School of Mathematics and Statistics, Xi’an Jiaotong University, Xi’an, China (e-mail: zhiqiang.wei@xjtu.edu.cn).

D. W. K. Ng is with the School of Electrical Engineering and Telecommunications, University of New South Wales, Sydney, NSW 2052, Australia (e-mail: w.k.ng@unsw.edu.au).
}% <-this % stops a space
  }
%   \IEEEauthorblockA{\IEEEauthorrefmark{1} Department of Electronic and Electrical Engineering, Southern University of Science and Technology, Shenzhen, China\\  \IEEEauthorrefmark{2} School of Mathematics and Statistics, Xi’an Jiaotong University, Xi’an, China  \\ \IEEEauthorrefmark{3} School of
% Electrical Engineering and Telecommunications, the University of New South Wales, Australia}
%   \IEEEauthorblockA{Email: wuj2021@mail.sustech.edu.cn, yuanwj@sustech.edu.cn, zhiqiang.wei@xjtu.edu.cn, zhangkc2022@mail.sustech.edu.cn, w.k.ng@unsw.edu.au}

\title{Low-Complexity Minimum BER Precoder Design for ISAC Systems: A Delay-Doppler Perspective}
\IEEEoverridecommandlockouts 
%\IEEEpubid{\makebox[\columnwidth]{Copyright Notice } 
%\hspace{\columnsep}\makebox[\columnwidth]{ }} 
\maketitle

\begin{abstract} 
Orthogonal time frequency space (OTFS) modulation is anticipated to be a promising candidate for supporting integrated sensing and communications (ISAC) systems, which is considered as a pivotal technique for realizing next-generation wireless networks. In this paper, we develop a minimum bit error rate (BER) precoder design for an OTFS-based ISAC system. In particular, the BER minimization problem takes into account the maximum available transmission power budget and the required sensing performance. \revise{Unlike previous studies that focused on ISAC in the time-frequency (TF) domain, we devise the precoder from the perspective of the delay-Doppler (DD) domain by exploiting the equivalent DD domain channel. The DD domain channel generally tends to be sparse and quasi-static, which is conducive to a low-complexity ISAC system design.} To address the non-convex optimization design problem,  we resort to optimizing the lower bound of the derived average BER by adopting Jensen's inequality. Subsequently, the formulated problem is decoupled into two independent sub-problems via singular value decomposition (SVD) methodology. We then theoretically analyze the feasibility conditions of the proposed problem and present a low-complexity iterative solution via leveraging the Lagrangian duality approach. Simulation results verify the effectiveness of our proposed precoder compared to the benchmark schemes and reveal the interplay between sensing and communication for dual-functional precoder design, indicating a trade-off where transmission efficiency is sacrificed for increasing transmission reliability and sensing accuracy.
\end{abstract}
\vspace{-0.4cm}
\section{Introduction }
\lettrine[lines=2]{E}{merging} 
  vehicular networks, regarded as a promising application of next-generation wireless systems, are expected to possess intelligent capabilities essential for autonomous driving and mitigating traffic congestion \cite{liang2018toward}. However, the realization of intelligent vehicular networks (IVNs) hinges prominently on two pivotal technologies, i.e., the functionalities of robust environmental sensing and reliable communication.  On the one hand, intelligent environmental sensing plays a critical role in enabling IVN participants to acquire information regarding road lanes, pedestrians, and unforeseen traffic accidents. Meanwhile, reliable communications empower the vehicles to seamlessly connect to the internet and share pertinent information, enabling the IVNs to perform timely decisions and conduct efficient resource scheduling. 

 \revise{Over the past few decades, sensing and communication systems have traditionally been developed independently, each assigned to orthogonal frequency bands. This separation has posed significant challenges for IVNs that rely on both sensing and communications functionalities to meet demanding requirements in localization, entertainment, as well as autonomous driving.}  Recently, a novel trend, where sensing and communications are co-designed in terms of hardware architecture and signal processing framework, has emerged, i.e., integrated sensing and communications (ISAC) \cite{9606831,10418473}. Through a unified design approach, ISAC enables a higher resource utilization efficiency and eventually facilitates mutual benefits. To this end, various advanced schemes have been proposed for realizing ISAC systems, involving joint resource scheduling \cite{10304366,10543024}, physical layer security optimization \cite{97su4,wujuav}, and frame structure design \cite{li2023frame}. In particular, the recent work \cite{10158322} considered a multi-UAV-enabled ISAC system, where UAVs are employed to track the ground user and send the downlink communication information. By leveraging real-time distance measurements, the UAV trajectory can be dynamically adapted, leading to a compelling performance. However, it is noteworthy that the consideration UAV of a single antenna contributes only modestly to sensing accuracy and overall throughput. Along this line, \cite{wei2023integrated} proposed an integrated sensing, navigation, and communication (ISNC) framework for
safeguarding multiple-input multiple-output (MIMO) UAV-enabled wireless networks against a mobile eavesdropper. Furthermore, a predictive beamforming design in vehicular networks adopting the ISAC technique was considered in \cite{9246715}, which has superior performance compared to traditional communication-only feedback-based schemes. 
To  theoretically analyze ISAC, the study in \cite{liu2023fundamental} summarized the performance metrics and bounds exploited in sensing and communications, which would provide useful insights and guidance for the development of 
ISAC for approaching the performance limits. One step further,
% the research in \cite{jiao2023rate} developed the information-theoretic limits of the bistatic ISAC system. This investigation entails a characterization of the capacity-distortion region through the exploration of diverse decoding-and-estimation methodologies.
in \cite{10X}, Xiong \textit{et al.} revealed a two-fold tradeoff in
ISAC systems, consisting of the subspace tradeoff (ST) and the
deterministic-random tradeoff (DRT) that depend on the resource
allocation and data modulation schemes employed for sensing and communications, respectively. 
% Building upon this framework,  the work presented in \cite{lu2023random} delved into sensing aspects utilizing random signaling within a MIMO system, in which a novel sensing performance metric, namely, ergodic linear minimum mean square error (ELMMSE), was proposed. The authors concluded that the data-dependent precoding scheme can attain optimized sensing performance at the price of high computational complexity. 

Despite considerable efforts being dedicated to the research of ISAC, it remains in its infancy. One of the primary challenges in realizing a practical ISAC system lies in the intricate task of designing a waveform that can flexibly balance the requirements of sensing and communications. Motivated by decades of research on orthogonal frequency division multiplexing (OFDM)-based communications, it is a natural evolution progression to devise ISAC systems by exploiting OFDM waveforms.
For example, the work of \cite{7970102} investigated a power allocation scheme in OFDM-based ISAC systems, where the 
mutual information (MI) between the random target impulse
response and the data information rate (DIR) of frequency-selective fading channel is considered. Besides, the authors in \cite{10247262} developed a weighted pilot optimization problem to minimize the bit error rate (BER) and average ranging measurement error. On the top of this work, \cite{10007638} devised a reference signal-based sensing scheme under the OFDM frame structure and power optimization for an energy-efficient ISAC system.
Although OFDM shows its superiority in robust synchronization and resilience against multi-path fading, it suffers from a high peak-to-average-power ratio (PAPR) and has severely degraded performance in high mobility scenarios due to the exceeding large Doppler spread and the induced inter-carrier interference. To circumvent these challenges, there is an emerging need for exploring alternative modulation schemes.

\revise{A novel paradigm, termed orthogonal time frequency space (OTFS) \cite{wei2022off}, has been introduced as a promising alternative to traditional modulation techniques. Unlike conventional OFDM, which modulates data symbols in the TF domain, OTFS operates in the delay-Doppler (DD) domain. This domain shift offers significant advantages, particularly in high-mobility scenarios, resulting in superior performance. The key benefit lies in the fact that fast time-varying channels exhibit quasi-static and sparse characteristics in the DD domain, in contrast to fluctuating and dense TF domain channel responses. This fundamental advantage has spurred explosive developments in research focused on advancing communications via OTFS signals.}
% In \cite{zhang}, a deep residual shrinkage network (DRSN) was proposed to learn the residual noise for recovering the channel information, and a self-adaptive threshold was employed to ensure the channel sparsity.
For example, the authors in \cite{wei} investigated the influence of transmitter and receiver windows on OTFS transmission and devised windows to enhance channel estimation and data detection performance. As a step forward, a Gaussian mixture distribution (GMD)-based approach was proposed to establish a flexible OTFS  detection framework \cite{10261989}. Meanwhile, various OTFS-related works focus on  radar sensing. For instance, the work of \cite{947hink} proposed a generalized likelihood ratio test-based detector under OTFS modulation. In particular, this work performs sensing tasks via the time-domain OTFS signal directly, rather than processing the signal in the DD domain, whose simulations demonstrate that the estimation performance is improved compared to the standard fast Fourier transform (FFT)-based OFDM radar. Furthermore, in \cite{10zhang}, the researchers proposed a two-step method to estimate the fractional delay and
Doppler shifts. Specifically, the authors first performed a two-dimensional (2D) correlation to obtain the integer parts, followed by a difference-based method to estimate the fractional parts of these parameters. \revise{Given the above OTFS-based sensing and communication works, it can be observed that the DD domain channel parameters naturally align with the physical characteristics of radio propagation environment \cite{10279816}. Moreover, OTFS demodulation process is exactly the same as the range-Doppler matrix computation process in radar sensing, indicating that the OTFS signal can be efficiently utilized for both sensing and communications tasks without different signal processing algorithms, which facilitates low computational complexity ISAC design. Consequently, OTFS has emerged as a promising candidate for realizing ISAC systems \cite{yuan2022orthogonal,10404096,10409528}.}

Inspired by the inherent advantages of OTFS, various interesting works, e.g., \cite{9724198, 9557830} have studied radar-assisted communication systems adopting OTFS modulation for both uplink and downlink transmissions in high-mobility vehicular networks. Thanks to the OTFS-ISAC signals, the roadside unit (RSU) can simultaneously transmit downlink messages to the vehicles and estimate the sensing parameters of vehicles based on the echoes. In this context, the RSU can efficiently predict the vehicles state in the following time slots. Consequently, it is unnecessary for the vehicles to perform explicit channel estimation before designing the precoding scheme or detecting the information. \revise{Existing OTFS-ISAC studies predominantly aimed to enhance spectral efficiency satisfying the required Cramér-Rao bound (CRB) constraint and thus typically addressed the trade-off between transmission efficiency and sensing performance. However, in scenarios with high mobility, communication reliability becomes a critical performance metric to maintain the robustness and stability of dynamic communication links.  In light of these emerging needs, a comprehensive OTFS-ISAC work, investigating the trade-off between communication reliability and sensing accuracy, has not yet been reported in the open literature. }
Furthermore, we note that most existing works on OTFS adopt complicated maximum-likelihood (ML) detection, introducing a relatively high complexity. To address this issue, in \cite{SHI2024104280}, the authors investigated an OTFS system model with zero-forcing (ZF) detectors that utilize ISAC to assist in designing the frame structure and analyze the channel estimation error and outage probability. However, these contributions mainly focus on signal processing in the TF domain, which does not fully exploit the characteristics of the DD domain.  As an alternative, the work of \cite{10138552} considered the predictive precoder design for ZF and minimum mean square error (MMSE) equalizers in the DD domain, which minimizes the BER for the OTFS-based ultra-reliable low-latency communication (URLLC) system. However, such a precoder, while effectively enhancing transmission reliability, lacks consideration of the sensing functionality. Consequently, it fails to explicitly address the critical sensing requirements, potentially limiting its direct applicability to an OTFS-ISAC system.

To the best of our knowledge, existing studies either utilize highly complicated detection algorithms for supporting OTFS-ISAC systems, e.g., \cite{9724198, 9557830}  or devise advanced precoders only for OTFS-based communication systems, e.g., \cite{10138552,9775701}. A comprehensive joint radar and communication precoder design, oriented to the DD domain, has not been reported in open literature yet. Motivated by this, in this paper, we devise the dual-functional precoder from the perspective of DD domain. In particular, the precoder is designed not only to maximize system reliability but also to guarantee sensing performance. The main contributions of this work can be summarized as follows:
\begin{itemize}
    \item In contrast to prior research that designs ISAC in the TF domain, we propose to adopt the OTFS signals for realizing ISAC systems and directly devise the precoder in the DD domain. To this end, we first derive an asymptotic BER expression by accounting for the precoder at the transmitter and the ZF equalizer at the receiver. Then, the dual functional precoder design problem is formulated to minimize the average BER performance in one OTFS block. Furthermore, the formulated problem concurrently takes into account the maximum available transmission power constraint and the required sensing performance accessed by the CRB. 
    \item To address the non-convex BER minimization problem, we establish a lower bound of BER by adopting Jensen's inequality at moderate-to-high SNRs, which is further exploited as the new objective function. Then, the optimization problem is further decoupled into two independent sub-problems by leveraging singular value decomposition (SVD). We theoretically analyze the feasibility conditions of the proposed problem and propose a computationally-efficient iterative solution by considering the Karush-Kuhn-Tucker (KKT) conditions and Lagrangian dual problem.
    \item  By solving the sub-problems separately, we achieve the optimal precoders by strategically allocating power across the eigen sub-channels with distinct gains in the DD domain. Extensive simulations demonstrate the advantages of our proposed approach compared to conventional ZF and MMSE schemes. We further show the non-trivial trade-off between sensing and communication for devising dual-functional precoder design.
\end{itemize}
The remainder of this paper is organized as follows. Section \ref{sm} introduces the OTFS modulation as well as the OTFS-ISAC system model in the DD domain. In Section \ref{sec3}, we formulate the BER minimization problem and the proposed solutions are derived in Section \ref{sec4}. Our simulation results are provided in Section \ref{sec5}, while Section \ref{sec6} concludes this paper.
 
\textit{Notations:} The boldface lowercase letter and boldface capital letter denote the vector and the matrix, respectively. The superscript $(\cdot)^{T}$ and $ (\cdot)^{H}$ denote the transposition and Hermitian operations, respectively. The left pseudo-inverse of $\mathbf{X}$ is denoted as $ \mathbf{X}^\dagger =(\mathbf{X}^H\mathbf{X})^{-1}\mathbf{X}^H$.  We adopt $|\cdot|$, $\mathbb{E}(\cdot)$, and tr($\cdot$) to denote the modulus of a complex variable, the expectation operation, and the trace operation, respectively;  $\otimes$, $\mathbf{F}_{{M}}$, and $\mathbf{I}_{M}$ represent the Kronecker product, the $M$-dimensional discrete Fourier transform (DFT) matrix and identity matrix, respectively; $\mathrm{vec}(\cdot)$ and $|| \cdot||$ denote the vectorization operator, the $2$-norm, respectively. The term $\mathcal{Q} $($ \cdot$) is defined by $\mathcal{Q}(x) \triangleq 1 / \sqrt{2 \pi} \int_x^{\infty} e^{-\lambda^2 / 2} d \lambda$.  We denote $\mathbb{C}$ and $\mathbb{R}$ as the set of complex and real numbers, respectively. $\mathbf{A}_{n,m}$ is the $(n,m)$-th entry of matrix $\mathbf{A}$.  The Gaussian distribution with mean $\boldsymbol{\mu}$ and covariance matrix $\mathbf{\Sigma}$ is denoted by $\mathcal{CN} (\boldsymbol{\mu}, \mathbf{\Sigma})$. $ \mathrm{Re}\left\{\cdot \right\} $ represents the real part of a complex variable. We adopt $\mathrm{diag}(\mathbf{a})$ to generate a diagonal matrix with $\mathbf{a}$ being the diagonal elements.

\section{System Model} \label{sm}
\begin{figure}[t]
    \centering
\captionsetup{font=small,name=Fig, labelsep=period,justification=raggedright,singlelinecheck=false}
\includegraphics[width=0.6\columnwidth]{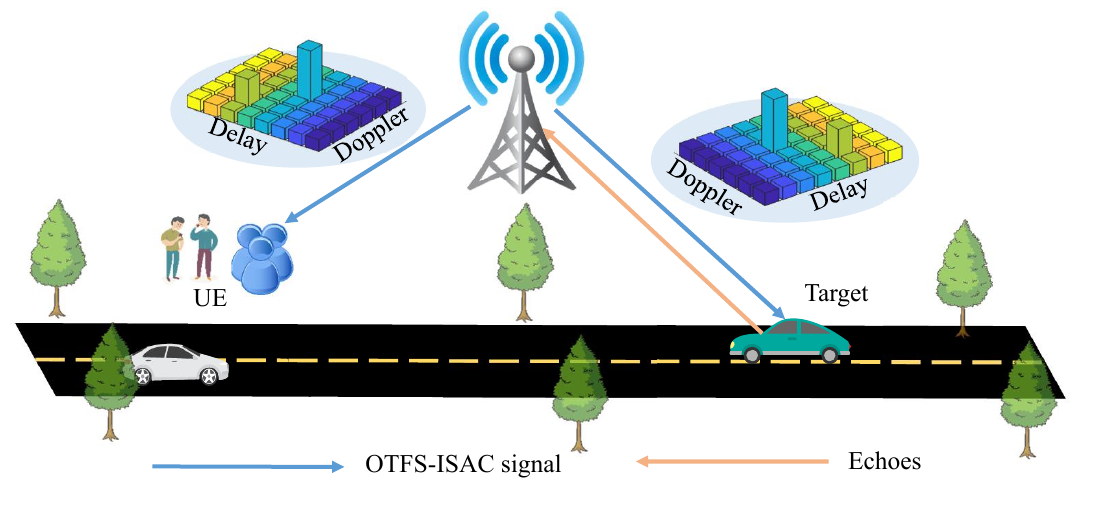}
    \caption{The considered OTFS-based ISAC scenario.}
    \label{sce} \vspace{-0.3cm}
\end{figure}

As illustrated in Fig. \ref{sce}, we consider an OTFS-based ISAC system, where a single-antenna transmitter sends the ISAC signal to execute the downlink communication with respect to (w.r.t) the user equipment (UE) and track the target simultaneously. In this section, we first briefly review the fundamentals of OTFS modulation as well as demodulation and then introduce our considered OTFS-enabled ISAC system model.
\begin{figure*}[t]
    \centering
\captionsetup{font=small,name=Fig,labelsep=period,justification=raggedright,singlelinecheck=false}
\includegraphics[width=\textwidth]{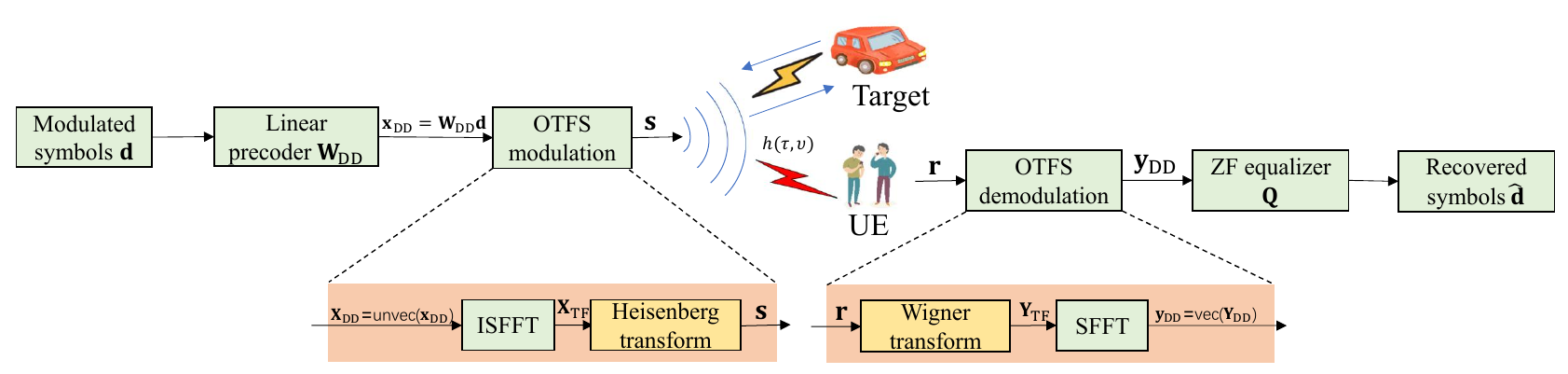}
    \caption{The adopted dual-functional precoder for OTFS-ISAC transmission.}
    \label{sce1} \vspace{-0.3cm}
\end{figure*}
\vspace{-0.4cm}
\subsection{OTFS Modulation and Demodulation} \label{sec1a}
Let $\mathbf{x}_\mathrm{DD} \in \mathbb{C}^{MN\times 1}$ be the transmitted ISAC symbol vector in the DD domain, where $M$ and $N$ represent the number of orthogonal subcarriers and the number of time slots for each OTFS frame, respectively. The vector $\mathbf{x}_\mathrm{DD}$ is arranged into a two-dimensional (2D) matrix, denoted by $\mathbf{X}_\mathrm{DD} \in \mathbb{C}^{M\times N}$, i.e., $\mathbf{x}_\mathrm{DD}=\mathrm{vec}(\mathbf{X}_\mathrm{DD})$. By denoting the subcarrier space and the symbol duration as $\Delta f$ and $T$,  the entire bandwidth and duration for each OTFS frame can be denoted by $M\Delta f$ and $NT$, respectively. In particular, the terms $\frac{1}{M\Delta f}$ and $\frac{1}{NT}$ are referred to as the delay resolution and Doppler resolution, respectively. \revise{ The entry in the $k$-th row and $l$-th column of $\mathbf{X}_\mathrm{DD}$ is represented by $X_\mathrm{DD}[k,l]$, in which $k\in \left\{ 0,\cdots,N-1\right\}$ is the Doppler index and $l\in \left\{ 0,\cdots,M-1\right\}$ is the delay index. }
 % The total bandwidth is $B=$ $M \Delta f$, and the total frame duration is $T_f=N T$. Furthermore, the OTFS system is critically sampled, i.e., $T \Delta f=1$.
 \revise{At the transmitter side, we first transform $X_\mathrm{DD}[k,l]$ into the TF domain signal $X_\mathrm{TF}[n,m]$ by adopting the inverse  symplectic finite Fourier transform (ISFFT) \cite{9508932, 10261989}, i.e.,
\begin{align}
    X_{\mathrm{TF}}[n, m]=\frac{1}{\sqrt{N M}} \sum_{k=0}^{N-1} \sum_{l=0}^{M-1} X_{\mathrm{DD}}[k, l] e^{j 2 \pi\left(\frac{n k}{N}-\frac{m l}{M}\right)}, \label{xtf}
\end{align}
where  $n\in \left\{ 0,\cdots,N-1\right\}$ and  $m\in \left\{ 0,\cdots,M-1\right\}$. } Each transmitted symbol in the TF domain is then transformed into the time domain signal $s(t)$ according to the OFDM modulation principle, which is given by
\begin{align}
    s(t)=\sum_{n=0}^{N-1} \sum_{m=0}^{M-1} X_{\mathrm{TF}}[n, m] g_\mathrm{tx}(t-n T) e^{j 2 \pi m \Delta f(t-n T)},
\end{align}
where $g_\mathrm{tx}(t)$ is the pulse shaping filter at the transmitter.  After passing through a time-varying channel, the transmitted signal can be received at the UE, which is expressed as  
\begin{align}
    r(t)=\iint h(\tau, \nu) e^{j 2 \pi \nu(t-\tau)} s(t-\tau) d \tau d \nu+z(t),
\end{align}
where $z(t)$ denotes the additive white Gaussian noise (AWGN) and $h(\tau, \nu)=\sum_{p=1}^P h_p \delta\left(\tau-\tau_p\right) \delta\left(\nu-\nu_p\right)$ represents the channel impulse response in the DD domain,
% \vspace{-0.17cm}
% \begin{align}
%     h(\tau, \nu)=\sum_{p=1}^P h_p \delta\left(\tau-\tau_p\right) \delta\left(\nu-\nu_p\right),\label{res}
% \end{align}
where $P$ represents the total number of resolvable independent propagation paths, $h_p$, $\tau_p \in\left[0, \tau_{\max }\right]$, and $\nu_p \in\left[-\nu_{\max }, \nu_{\max }\right]$ are the complex path gain, the delay, and Doppler shift associated with the $p$-th path, respectively. The terms $\tau_{\max }$ and $\nu_{\max }$ denote the maximum delay and Doppler shift, respectively. 
\revise{
 At the receiver side, let us denote $A_{g_{\mathrm{rx}}, r}(f, t)$ as the matched filter output, which is given by 
 \begin{align}
     A_{g_{\mathrm{rx}}, r}(f, t) \triangleq \int r\left(t^{\prime}\right) g_{\mathrm{rx}}^*\left(t^{\prime}-t\right) e^{-j 2 \pi f\left(t^{\prime}-t\right)} d t^{\prime},\label{cross}
 \end{align}
 where $g_\text{rx}(t)$ is the receiving filter.
Then, the received signal in the TF domain can be obtained by sampling $A_{g_{\mathrm{rx}}, r}(f, t)$, i.e., $Y_\mathrm{TF} [n,m]=A_{g_{\mathrm{rx}}, r}(f, t)|_{f=m\Delta f, t=nT}.$ 
% \begin{align}
%    Y_\mathrm{TF} [n,m]=A_{g_{\mathrm{rx}}, r}(f, t)|_{f=m\Delta f, t=nT}.
% \end{align}
Subsequently, the SFFT is adopted to convert the received signal from the TF domain into the DD domain, which is  expressed as 
{\small \begin{align}
    Y_{\mathrm{DD}}[k, l]=\frac{1}{\sqrt{N M}} \sum_{n=0}^{N-1} \sum_{m=0}^{M-1} Y_{\mathrm{TF}}[n, m] e^{-j 2 \pi\left(\frac{n k}{N}-\frac{m l}{M}\right)}+\Bar{z}[k,l],
 \end{align}}
in which $\Bar{z}[k,l]$ denotes the noise sample in the DD domain associated with the term $Y_{\mathrm{DD}}[k, l]$.} Similar to $X_\mathrm{DD}[k,l]$, $Y_\mathrm{DD}[k,l]$ can also be arranged into a 2D matrix, denoted by $\mathbf{Y}_\mathrm{DD}$. 

\revise{To facilitate the precoder design, we re-express the OTFS modulation and demodulation in a compact matrix form.} According to (\ref{xtf}), the transmitted OTFS-ISAC symbol matrix in the TF domain, say $\mathbf{X}_\mathrm{TF}$, can be obtained by applying ISFFT to $\mathbf{X}_\mathrm{DD}$.  Thus, $\mathbf{X}_{\mathrm{TF}}$ can be written as $ \mathbf{X}_{\mathrm{TF}}=\mathbf{F}_{{M}} \mathbf{X}_{\mathrm{DD}} \mathbf{F}_{N}^{{H}}.$
% \begin{align}
%     \mathbf{X}_{\mathrm{TF}}=\mathbf{F}_{{M}} \mathbf{X}_{\mathrm{DD}} \mathbf{F}_{N}^{{H}}.
% \end{align}
Assuming that we adopt a rectangular pulse shaping \cite{9508932} for both transmitter and receiver. Thus, we have the transmitted time-domain signal $\mathbf{s}=\operatorname{vec}\left(\mathbf{F}_{{M}}^{{H}} \mathbf{X}_{\mathrm{TF}}\right) \in \mathbb{C}^{M N \times 1}$ by (discrete) Heisenberg transform.
% \begin{align}
% \mathbf{s}=\operatorname{vec}\left(\mathbf{F}_{{M}}^{{H}} \mathbf{X}_{\mathrm{TF}}\right).
% \end{align}
At the receiver, the signal can be straightforwardly given by $\mathbf{r}=\mathbf{H}_\mathrm{T} \mathbf{s}+\mathbf{n}$,
% \begin{align}
% \mathbf{r}=\mathbf{H}_\mathrm{T} \mathbf{s}+\mathbf{n},
% \end{align}
where $\mathbf{n} $ is the AWGN and $\mathbf{H}_\mathrm{T}$ represents the time-domain channel matrix, which is expressed as 
\vspace{-0.17cm}
\begin{align}
    \mathbf{H}_\mathrm{T}=\sum_{p=1}^P h_p' \boldsymbol{\Delta}^{k_p}\boldsymbol{\Pi}^{l_p} , \label{TIMe}
\end{align}
where $h_p'=h_pe^\frac{-j2\pi k_pl_p}{MN}$ with  $l_p$ and $k_p$ being the delay and Doppler-shift taps corresponding to the $p$-th path, which can be derived by $l_p=M \Delta f \tau_p$ and $k_p=N T \nu_p$, respectively. In practice, the sampling time $1/M\Delta f$ should be sufficiently small, and hence, $l_p$ can be assumed as an integer for ease of exposition \cite{9724198,10138552}. Conversely, $k_p$ should encompass both the integer and fractional components. In (\ref{TIMe}),  $\mathbf{\Pi}$ denotes the forward cyclic shift matrix given by 
\begin{align}
\boldsymbol{\Pi}=\left[\begin{array}{cccc}
0 & \cdots & 0 & 1 \\
1 & \ddots & 0 & 0 \\
\vdots & \ddots & \ddots & \vdots \\
0 & \cdots & 1 & 0
\end{array}\right] \in \mathbb{R}^{M N \times M N},
\end{align}
and $\mathbf{\Delta}^{k_p}=\operatorname{diag}\left(1, e^{j 2 \pi k_p \frac{1}{M N}}, \ldots, e^{j 2 \pi k_p \frac{M N-1}{M N}}\right) \in \mathbb{C}^{M N \times M N}$ is the diagonal Doppler matrix.
% \begin{align}
%     \mathbf{\Delta}^{k_p}= \operatorname{diag}\left(e^{\frac{j 2 \pi k_p \times 0}{M N}}, e^{\frac{j 2 \pi k_p \times 1}{M N}}, \ldots, e^{\frac{j 2 \pi k_p \times(M N-1)}{M N}}\right).
% \end{align}
% \begin{align}
%     \mathbf{\Delta}^{k_p}=\left[\begin{array}{ccc}
% e^{\frac{j 2 \pi k_p(0)}{M N}} & 0 & 0 \\
% 0 & \ddots & 0 \\
% 0 & 0 & e^{\frac{j 2 \pi k_p(M N-1)}{M N}}
% \end{array}\right]
% \end{align}
 With the Wigner transform and the SFFT in hand, it is evident that the output symbol matrix in the DD domain can be written as $\mathbf{Y}_\mathrm{DD}=\mathbf{R}_\mathrm{T} \mathbf{F}_{{N}},$ 
% \begin{align}
% \mathbf{Y}_\mathrm{DD}=\mathbf{R}_\mathrm{T} \mathbf{F}_{{N}},
% \end{align}
where $\mathbf{R}_\mathrm{T} $ is obtained by arranging $\mathbf{r}$ into a 2D matrix. Then, we can summarize the input-output relationships in the DD domain, which can be written as
\begin{align} \mathbf{y}_\mathrm{DD}  
 =\left(\mathbf{F}_{{N}} \otimes \mathbf{I}_{{M}}\right) \mathbf{H}_\mathrm{T}\left(\mathbf{F}_{{N}}^{{H}} \otimes \mathbf{I}_{{M}}\right) \mathbf{x}_\mathrm{DD}+\left(\mathbf{F}_{{N}} \otimes \mathbf{I}_{{M}}\right) \mathbf{n},\label{sum}
\end{align}
where $=\operatorname{vec}(\mathbf{Y}_\mathrm{DD})$. Furthermore, by denoting the equivalent channel matrix size of $MN\times MN$ in the DD domain as $\mathbf{H}_{\mathrm{DD}} \triangleq$ $\left(\mathbf{F}_{{N}} \otimes \mathbf{I}_{{M}}\right) \mathbf{H}_\mathrm{T}\left(\mathbf{F}_{{N}}^{{H}} \otimes \mathbf{I}_{{M}}\right)$, (\ref{sum}) can be simplified as 
\begin{align}
\mathbf{y}_\mathrm{DD}=\mathbf{H}_{\mathrm{DD}} \mathbf{x}_\mathrm{DD}+\left(\mathbf{F}_{{N}} \otimes \mathbf{I}_{{M}}\right) \mathbf{n}. \label{ddchan}
\end{align}% In (\ref{ddchan}), it is noted that the DD domain channel response is a superposition of $P$ paths. In high-mobility scenarios, the superposed DD channel response is still sparse and resolvable, resulting in better sensing and communication performance compared to OFDM \cite{9508932}. 
\subsection{OTFS-Enabled ISAC System Model}
\revise{Without loss of generality, we employ a linear precoder in the DD domain prior to implementing OTFS transmission as shown in Fig. \ref{sce1}. The 
 transmitted symbols $\mathbf{x}_\mathrm{DD}$ are generated by linearly mapping the data symbols $\mathbf{d}$ using a precoder $\mathbf{W}_\mathrm{DD}$, i.e.,  $\mathbf{x}_\mathrm{DD}=\mathbf{W}_\mathrm{DD}\mathbf{d}$.  Here, $\mathbf{W}_\mathrm{DD}\in \mathbb{C}^{MN\times MN} $ is the dual-functional precoder matrix to be designed and $\mathbf{d}\in \mathbb{C}^{MN\times 1}$ represents the $MN$ data symbols drawn from $\mathbb{A}$}\footnote{\revise{ In this work, our primary focus is on the vector \(\mathbf{d}\) of dimension \(MN\), i.e., the full eigen sub-channels mode \cite{9775701}. However, when the number of transmitted symbols is less than \(MN\), corresponding to the eigen sub-channels dropping mode \cite{10138552}, the design of low-complexity precoders may necessitate further consideration, which remains for our future work.}}, \revise{which is a set of discrete modulation alphabets, e.g., quadrature amplitude modulation (QAM).
% \footnote{Since the precoder $\mathbf{W}_\mathrm{DD}$ is required to linearly transform $MN$ data symbols, it should be with a full rank, i.e., {rank}$(\mathbf{W}_\mathrm{DD})=MN$.}. 
To facilitate the solution design, we assume that the information vector $\mathbf{d}$ is with unit power, resulting in $\mathbb{E}\left\{\mathbf{dd}^H\right\}=\mathbf{I}_{MN}$. As a result, the entire transmission power becomes 
\begin{align}
\mathrm{tr}\left(\mathbb{E}\left\{\mathbf{W}_\mathrm{DD}\mathbf{d}\left(\mathbf{W}_\mathrm{DD}\mathbf{d}\right)^{{H}}\right\}\right)=\mathrm{tr}\left(\mathbf{W}_\mathrm{DD}\mathbf{W}_\mathrm{DD}^{{H}}\right).
\end{align}
By transmitting the OTFS-ISAC signals to the UE, the received signals in the DD domain can be expressed as
\begin{align}
\mathbf{y}_\mathrm{c}=\mathbf{H}_{\mathrm{c,DD}} \mathbf{W}_\mathrm{DD}\mathbf{d}+ \mathbf{n}_\mathrm{c}, \label{coms}
\end{align}
where $\mathbf{H}_{\mathrm{c,DD}}$, defined similarly to $\mathbf{H}_\mathrm{DD}$ in Sec. \ref{sec1a}, is the equivalent DD domain channel from the transmitter to the UE and $\mathbf{n}_c $ is the AWGN vector, which follows $\mathbf{n}_c \sim \mathcal{CN}(\mathbf{0},\sigma_c^2\mathbf{I}_{MN})$ with $\sigma_c^2$ being the noise power}\footnote{\revise{In practice, channel state information (CSI) may be imperfect. Against this background, robust designs are critical for ensuring reliable system performance in real-world applications. Since our primary focus is to explore the trade-off between communication reliability and sensing accuracy, we adopt a common assumption of perfect CSI in this work, as considered in the literature \cite{10279816,9557830,9724198}. }}.
Meanwhile, the ISAC waveform is also radiated and reflected by the target for tracking purpose. The reflected echo at the transmitter is given by 
 \begin{align}
\mathbf{y}_\mathrm{s}=\mathbf{H}_\mathrm{s,DD}\mathbf{x}_\mathrm{DD}+\mathbf{n}_\mathrm{s}, \label{sens}
\end{align}
where $\mathbf{H}_\mathrm{s,DD}$ is the target response matrix in the DD domain and $\mathbf{n}_s$ is the AWGN obeying $\mathbf{n}_\mathrm{s}\sim \mathcal{CN}(\mathbf{0},\sigma_s^2\mathbf{I}_{MN})$, where  $\sigma_s^2$ is the noise power.
\vspace{-0.3cm}
% with $\mathbf{R}_{MN}=\sigma_s^2\mathbf{I}_{MN}$.
% In general, it is clear that a radar system is capable of performing both search and tracking tasks. The search mode is referred to as scanning the unknown environments to search for potential targets of interest. In this case, there may exist multiple scatters, resulting in that the ISAC signal is transmitted over multiple paths. For the tracking mode, the radar focuses on the specific target's location and velocity as it moves in the radar's coverage region, in which case we assume is transmitted over a signal path.In practice, a radar system possesses the capability to execute both search and tracking functions. The search mode entails the exploration of unknown environments to identify potential targets of interest. In this phase, it is convinced that multiple scattering phenomena may occur, resulting in the ISAC signal being transmitted over multi-path. On the other hand, for the tracking mode, the radar focuses on the specific target's location and velocity as it moves in the radar's coverage area. In this scenario, one could reasonably assume that the ISAC waveform is transmitted via one signal path. Without loss of generality, we will discuss the dual-functional precoder design for both the single path case and multiple paths case in Sec. \ref{sec_solution}. 
\section{Problem Formulation} \label{sec3}
In this section, we first develop both the sensing and communication metrics, building upon the introduced model in Sec. \ref{sm}. These metrics are adopted as the criteria to formulate our considered precoder design problem.  At the UE, the received OTFS-ISAC symbols, denoted by $\hat{\mathbf{d}}$, can be recovered as
\vspace{-0.3cm}
\begin{align}
\mathbf{\hat{d}}=\mathbf{QH}_{\mathrm{c,DD}}\mathbf{W}_\mathrm{DD}\mathbf{d}+\mathbf{Q}\mathbf{n}_\mathrm{c},
\end{align}
where $\mathbf{Q}\in \mathbb{C}^{MN\times MN}$ denotes the linear equalizer. For ease of exposition,  we exploit the widely-adopted ZF equalizer $\mathbf{Q}=\left(\mathbf{H}_{\mathrm{c,DD}}\mathbf{W}_\mathrm{DD} \right)^\dagger$. 
% \begin{align}
% \mathbf{Q}=\left(\mathbf{H}_{\mathrm{c,DD}}\mathbf{W}_\mathrm{DD} \right)^\dagger.    
% \end{align}
% \begin{align}
% \mathbf{Q}=\left(\mathbf{W}_\mathrm{DD}^{H}\mathbf{H}_{\mathrm{c,DD}}^{{H}}\mathbf{H}_{\mathrm{c,DD}}\mathbf{W}_\mathrm{DD}\right)^{-1}\mathbf{W}_\mathrm{DD}^{{H}}\mathbf{H}_{\mathrm{c,DD}}^{{H}}.
% \end{align}
It is noted that $\mathbf{H}_{\mathrm{c,DD}}$ is randomly generated and can be almost guaranteed to be non-singular. The mean square error (MSE) of the $i$-th received symbol ($i\in [1,MN]$) is given by the $i$-th diagonal element of the MSE matrix $\mathbf{E}_\mathbf{\hat{d}} $ \cite{palomar2005minimum}, i.e., $\mathrm{MSE}_i=\left[\mathbf{E}_\mathbf{\hat{d}} \right]_{i,i}$, where $\mathbf{E}_\mathbf{\hat{d}} $ is defined as\footnote{ Following \cite{palomar2005minimum}, the MSE matrix can be given by $\mathbf{E}_\mathbf{\hat{d}}= \left(\mathbf{A}-\mathbf{I}_{MN}\right)\left(\mathbf{A}-\mathbf{I}_{MN}\right)^H +\sigma_c^2\mathbf{Q}  \mathbf{Q}^H$ with $\mathbf{A}=\mathbf{QH}_{\mathrm{c,DD}}\mathbf{W}_\mathrm{DD}$. Due to the ZF constraint,  we have $\mathbf{A}=\mathbf{QH}_\mathrm{c,DD}\mathbf{W}_\mathrm{DD}=\mathbf{I}_{MN}$, such that the MSE matrix is reduced to (\ref{msed}). } 
\begin{align}
 \mathbf{E}_\mathbf{\hat{d}}=\mathbb{E}\left[ \left(\hat{\mathbf{d}}-\mathbf{d} \right)\left(\hat{\mathbf{d}}-\mathbf{d} \right)^H\right]= \sigma_c^2\mathbf{Q}  \mathbf{Q}^H. \label{msed}
    % \left(\mathbf{Q}^H \mathbf{H}_\mathrm{c,DD} \mathbf{W}_\mathrm{DD}-\mathbf{I}\right)\left(\mathbf{W}_\mathrm{DD}^H \mathbf{H}_\mathrm{c,DD}^H \mathbf{Q}-\mathbf{I}\right)+
\end{align}
The ZF receiver is also optimal in the sense of maximizing the signal-to-interference-plus-noise ratios (SINRs) of all received symbols, which are directly associated with the MSEs by $\operatorname{SINR}_{i}=1/{\operatorname{MSE}_{i}}$, \cite{palomar2005minimum,9775701}, 
% \begin{align}
%     \operatorname{SINR}_{i}=\frac{1}{\operatorname{MSE}_{i}}, \label{lik}
% \end{align}
where $\operatorname{SINR}_{i}$ is the SINR of the $i$-th received symbol.  Consequently,  the SINR of the $i$-th symbol can be re-expressed as 
\begin{align}
\mathrm{SINR}_i =\frac{1}{\left[\sigma_c^2\left(\mathbf{W}_\mathrm{DD}^{{H}}\mathbf{H}_{\mathrm{c,DD}}^{{H}}\mathbf{H}_{\mathrm{c,DD}}\mathbf{W}_\mathrm{DD}\right)^{-1}\right]_{i,i}}.
\end{align}
\revise{With the assumption of Gray encoding \cite{ding2003minimum}, the BER of the $i$-th symbol in the scenario of $M_{\mathrm{Mod}}$-ary QAM constellation can be asymptotically characterized as $\mathrm{BER}_i\approx \alpha \mathcal{Q}\left(\sqrt{\beta \mathrm{SINR}_i}\right)$\footnote{\revise{The commonly adopted $\mathcal{Q}(\cdot)$ function and the complementary error function
$ \mathrm{erfc} (\cdot)$ are directly linked by $\mathcal{Q}(x)=\frac{1}{2}\mathrm{erfc} (
\frac{x}{\sqrt{2}})$.}},
% \begin{align}
% \mathrm{BER}_i\approx \alpha \mathrm{erfc}\left(\sqrt{\beta \mathrm{SINR}_i}\right),
% \end{align}
% \begin{align}
%     \mathrm{BER}_m\approx \alpha \mathrm{erfc}\left(\sqrt{\beta \mathrm{SINR}_m}\right),
% \end{align}
in which $\alpha=(4-4/\sqrt{M_{\mathrm{Mod}}})/\log_2M_{\mathrm{Mod}}$ and $\beta=3/(M_{\mathrm{Mod}}-1)$ \cite{9775701}. We now utilize the average BER of the detected symbols to evaluate the reliability performance of each received OTFS block, which is given by
\vspace{-0.24cm}
 { \small \begin{align}
\nonumber P_e &=\frac{1}{MN} \sum_{i=1}^{MN} \mathrm{BER}_i \\ &=\frac{\alpha}{MN} \sum_{i=1}^{MN} \mathcal{Q}\left(\sqrt{\frac{\beta}{\sigma_c^2\left[\left(\mathbf{W}_\mathrm{DD}^{{H}} \mathbf{H}_{\mathrm{c,DD}}^{{H}} \mathbf{H}_{\mathrm{c,DD}} \mathbf{W}_\mathrm{DD}\right)^{-1}\right]_{i,i}}}\right). \label{ber} \vspace{-0.3cm}
\end{align}}}
% \begin{align*}
% & \geq \alpha \operatorname{erfc}\left(\sqrt{\frac{\beta K}{\sigma^2 \operatorname{tr}\left(\left(\kappa\sigma^2 \mathbf{I}+\mathbf{W}^{\mathbf{H}} \mathbf{H}_{\mathrm{c,dd}}^{\mathbf{H}} \mathbf{H}_{\mathrm{c,dd}} \mathbf{W}\right)^{-1}\right)}-\beta \kappa}\right) \\
% & \triangleq P_{e, \mathrm{LB}} 
% \end{align*}

We then investigate the metric for evaluating the tracking purpose. Since we mainly focus on a tracking scenario, it is assumed that the signal propagation between the transmitter and target is dominated by the line-of-sight (LoS) path, i.e., $P=1$. In practice, it is imperative to recognize the presence of undesired echoes resulting from reflections off terrestrial objects. Extensive investigation and utilization of filtering and deconvolutional techniques have been undertaken to effectively mitigate clutter interference \cite{xu2015range,7862855}. \revise{Notice that the delay indices are already assumed to be integers in Sec. \ref{sec1a}, which further supports a reasonable assumption that the delay estimation is sufficiently accurate, as commonly adopted in the literature \cite{10zhang,9456029}. This assumption allows us to concentrate on improving the Doppler estimation performance.  In light of these prior studies, this work is dedicated to the analysis and optimization of Doppler frequency shift estimation}.  To access the estimation performance, the MSE has been frequently adopted, which, however, usually cannot be derived in a closed-form expression and thus may cause an intractable optimization problem. To circumvent this challenge, we adopt the CRB as the sensing performance metric, which is capable of providing a lower bound for unbiased parameter estimators. Based on CRB theorem \cite{kay1993fundamentals}, the variance of unbiased estimation $\tilde{\nu}$ is required to satisfy $\mathrm{var}(\tilde{\nu})\ge \frac{1}{I(\nu)}$, where ${I(\nu)}$ is the Fisher information given by
% \begin{align}
%     \mathrm{var}(\tilde{\nu})\ge \frac{1}{-\mathbb{E}\left[ \frac{\partial \ln p(\mathbf{y}_{\mathrm{s}};\nu) }{\partial \nu^2} \right]},
% \end{align}
%  where $p(\mathbf{y}_{\mathrm{s}};\nu)$ is the likelihood function of $\mathbf{y}_{\mathrm{s}}$ conditioned on $\nu$ and the entire term in the denominator is the Fisher information.
\begin{align}
  I(\nu)=  \frac{1}{2}\operatorname{tr}\left(\mathbf{R}_{MN}^{-1} \frac{\partial \mathbf{R}_{MN}}{\partial \nu} \mathbf{R}_{MN}^{-1} \frac{\partial \mathbf{R}_{MN}}{\partial \nu}\right)+ \operatorname{Re}\left\{\frac{\partial {\mathbf{y}_\mathrm{s}}^{{H}}}{\partial \nu} \mathbf{R}_{MN}^{-1} \frac{\partial {\mathbf{y}_s}}{\partial \nu}\right\},\label{CRLB}
\end{align}
with $\mathbf{R}_{MN}=\sigma_s^2\mathbf{I}_{MN}$.
% \begin{align}
%     {\mathbf{J}}_{i_1, j_1}=\mathbb{E}\left[\frac{\partial\ln p(\mathbf{y}_\mathrm{s};\boldsymbol{\nu})}{\partial\nu_{i_1}}\frac{\partial\ln p(\mathbf{y}_\mathrm{s};\boldsymbol{\nu})}{\partial\nu_{j_1}}\right],  i_1, j_1 \in\{1,2,\cdots,P\},
% \end{align}
% since the noise $\mathbf{n}$ is independent to the channel, thus the FIM is dominated by the last term.  We denote $\dot{\mathbf{H}}_{s,dd,i}=\frac{\partial \mathbf{H}_{s,dd}}{\partial \boldsymbol{\nu_i}}$ ,  then
% \begin{align}
%     {\mathbf{J}}_{i, j}=\mathbb{E}\left[\frac{\partial\ln p(\mathbf{y};\boldsymbol{\nu})}{\partial\nu_i}\frac{\partial\ln p(\mathbf{y};\boldsymbol{\nu})}{\partial\nu_j}\right] \\
% =\mathbb{E} \left[ {\frac{\partial \left(\mathbf{H_{s,dd}x}\right)}{\partial\nu_{i}}}^{H}\mathbf{R_n}^{-1}[\mathbf{y}\mathbf{y}^{H}]\mathbf{R_n}^{-1}{\frac{\partial\left(\mathbf{H_{s,dd}x}\right)}{\partial\nu_{j}}}\right] \\
%  \quad = \mathbb{E} \left[ \mathbf{x^H\dot{H}_{s,dd,i}^HR_n^{-1}\dot{H}_{s,dd,j}x} \right] \\
%  \qquad \quad=\mathbb{E}  \left[ \mathrm{tr}\left(\mathbf{x^H\dot{H}_{s,dd,i}^HR_n^{-1}\dot{H}_{s,dd,j}x} \right) \right] \\ 
% \qquad \qquad  \qquad  \quad\qquad =\frac{ 1}{\sigma_{\mathrm{R}}^2} \operatorname{tr}\left(\dot{\mathbf{H}}_{s,dd,i} \mathbf{FF^H}\dot{\mathbf{H}}_{s,dd,j}^{\mathrm{H}}\right),  i, j \in\{1,2,\cdots, p\}
% \end{align}
 By simple algebraic derivation, we can rewrite (\ref{CRLB}) as $ I(\nu)=\frac{1}{\sigma_{{s}}^2} \operatorname{tr}\left(\dot{\mathbf{H}}_{\mathrm{s,DD}} \mathbf{W}_\mathrm{DD}\mathbf{W}_\mathrm{DD}^H\dot{\mathbf{H}}_{\mathrm{s,DD}}^{{H}}\right),$
 % \begin{align}
 %     I(\nu)=\frac{1}{\sigma_{{s}}^2} \operatorname{tr}\left(\dot{\mathbf{H}}_{\mathrm{s,DD}} \mathbf{W}_\mathrm{DD}\mathbf{W}_\mathrm{DD}^H\dot{\mathbf{H}}_{\mathrm{s,DD}}^{{H}}\right),  
 % \end{align}
where $\dot{\mathbf{H}}_{\mathrm{s,DD}}=\frac{\partial {\mathbf{H}_\mathrm{s,DD}}}{\partial \nu}$ is given by
\begin{align}
\dot{\mathbf{H}}_{\mathrm{s,DD}} =h_{s}\left(\mathbf{F}_N \otimes \mathbf{I}_M\right) \mathbf{D}_{\nu }\mathbf{\Delta}^{k_s}\mathbf{\Pi}^{l_s} \left(\mathbf{F}_N^{\mathrm{H}} \otimes \mathbf{I}_M\right), \label{gra}
\end{align}
with  
$
\mathbf{D}_{\nu}=\operatorname{diag}\left(j \frac{2 \pi T}{M}[-l_{s}, \ldots, M N-1-l_{s}]^{\mathrm{T}}\right) .
$
In (\ref{gra}), the definitions of $h_{s}$, $l_{s}$, and $k_{s}$ are similar to their counterparts in (\ref{TIMe}), respectively.
% Based on the above discussion, we now have the complete FIM w.r.t the Dopplers of the $P$ paths, which is available in (\ref{fim}) on the top of the next page.
% \begin{figure*}[t]
%     \begin{align}
% \mathbf{J}=\left[ \begin{array} {cccc}
% \frac{1}{\sigma_{{s}}^2} \operatorname{tr}\left(\dot{\mathbf{H}}_{\mathrm{s,DD},1} \mathbf{W}_\mathrm{DD}\mathbf{W}_\mathrm{DD}^H\dot{\mathbf{H}}_{\mathrm{s,DD},1}^{{H}}\right) & \cdots & \cdots&  \frac{1}{\sigma_{{s}}^2} \operatorname{tr}\left(\dot{\mathbf{H}}_{\mathrm{s,DD},1} \mathbf{W}_\mathrm{DD}\mathbf{W}_\mathrm{DD}^H\dot{\mathbf{H}}_{\mathrm{s,DD},P}^{{H}}\right) \\
% \vdots & \ddots & &\vdots   \\
% \vdots &  &\ddots &\vdots\\
% \frac{1}{\sigma_{{s}}^2} \operatorname{tr}\left(\dot{\mathbf{H}}_{\mathrm{s,DD},P} \mathbf{W}_\mathrm{DD}\mathbf{W}_\mathrm{DD}^H\dot{\mathbf{H}}_{\mathrm{s,DD},1}^{{H}}\right) &  \cdots &\cdots& \frac{1}{\sigma_{{s}}^2} \operatorname{tr}\left(\dot{\mathbf{H}}_{\mathrm{s,DD},P} \mathbf{W}_\mathrm{DD}\mathbf{W}_\mathrm{DD}^H\dot{\mathbf{H}}_{\mathrm{s,DD},P}^{{H}}\right)
% \end{array} \right] \
% \label{fim}
% \end{align}
% \hrulefill
% \end{figure*}
Consequently, the CRB, being the inverse of Fisher information, is expressed as   
% \begin{align}
%     \mathrm{CRLB}= \mathrm{FIM}^{-1}.
% \end{align}
% Assume that there only exists one single propagation path in our considered tracking scenario, the CRLB is reduced as
\begin{align}
     {\mathrm{CRB}}=\left(\frac{1}{\sigma_{{s}}^2} \operatorname{tr}\left(\dot{\mathbf{H}}_\mathrm{s,DD} \mathbf{W}_\mathrm{DD}\mathbf{W}_\mathrm{DD}^{H}\dot{\mathbf{H}}_\mathrm{s,DD}^{{H}}\right)\right)^{-1}. \label{spCRLB}
\end{align}
The DD precoder design problem aims to improve the BER performance, which concurrently is constrained by the maximum available transmission power and the required CRB threshold. To this end, the BER minimizing problem can be formulated as 
\begin{subequations} \label{pro}
\begin{align} 
 \min_{\mathbf{W}_\mathrm{DD}} \quad& P_e \\
\operatorname{s.t.} \quad & \mathrm{CRB} \le \gamma_c, \ \operatorname{tr} (\mathbf{W}_\mathrm{DD}\mathbf{W}_\mathrm{DD}^H)\le P_T, \label{pe}
\end{align}
\end{subequations}
where $\gamma_c$ is the required maximum tolerable CRB threshold and $P_T$ is the total transmission power budget. It is noted that problem (\ref{pro}) is difficult to solve by conventional convex optimization methods since the objective function is discrete and non-convex w.r.t $\mathbf{W}_\mathrm{DD}$. In the following, we address the non-convex challenge and derive a sub-optimal solution to problem (\ref{pro}).
\section{DD Precoder Optimization} 
\label{sec4}
In this section, we propose an efficient algorithm to solve problem (\ref{pro}). In particular, we first derive a lower bound of $P_e$  and reformulate a new optimization problem to approximate the original problem (\ref{pro})\footnote{ In general, we minimize the upper bound of the objective function as an approximation since the lower bound of a minimization problem may be unattainable. However, in this work, minimizing the lower bound of $P_e$ holds significance as it can be achieved through the utilization of our proposed method, which we will prove in Sec. \ref{iv-b}.}. Furthermore, the SVD method is employed to tackle the complicated constraints. Then, we analyze the feasibility conditions of the proposed problem, followed by presenting the optimal DD precoder design.
\vspace{-0.4cm}
\subsection{Problem Reformulation}
\revise{To convexify the objective function of (\ref{pro}), let us define $f(x)=\mathcal{Q}\left( 1/{\sqrt{\frac{\sigma_c^2}{\beta}x}}\right)$, where $x>0$. Then, by calculating the first derivative and second derivative, we have 
\begin{align}
    \frac{df}{dx}&= \frac{1}{2}   \sqrt{\frac{\beta}{2\pi \sigma_c^2}} \exp \left(-\frac{\beta}{ 2\sigma_c^2 x}\right) x^{-\frac{3}{2}}\quad \text{and}\\
    \frac{d^2f}{dx^2}&=\frac{1}{2}  \sqrt{\frac{\beta}{2\pi \sigma_c^2}} \exp \left(-\frac{\beta}{ 2\sigma_c^2 x}\right) \left( \frac{\beta}{2\sigma_c^2x} -\frac{3}{2}\right) x^{-\frac{5}{2}}.
\end{align}
It can be observed that $f(x)$ is convex if and only if the inequality $ \frac{d^2f}{dx^2}\ge0$ holds, i.e., $x\le\frac{\beta}{3\sigma_c^2}$. Employing this fact to (\ref{ber}), we can conclude that $f\left(\left[\left(\mathbf{W}_\mathrm{DD}^{{H}} \mathbf{H}_{\mathrm{c,DD}}^{{H}} \mathbf{H}_{\mathrm{c,DD}} \mathbf{W}_\mathrm{DD}\right)^{-1}\right]_{i,i}\right)$ is convex under moderate-to-high SNRs \cite{9775701}, i.e., 
\begin{align}
    \sigma_c^2\le\beta/\left(3 \left[\left(\mathbf{W}_\mathrm{DD}^{{H}} \mathbf{H}_{\mathrm{c,DD}}^{{H}} \mathbf{H}_{\mathrm{c,DD}} \mathbf{W}_\mathrm{DD}\right)^{-1}\right]_{i,i}\right), \forall i. \label{high}
\end{align} 
With this mild condition, a lower bound of (\ref{ber}) can be established by adopting Jensen's inequality, i.e.,
 \begin{align}
  \nonumber P_e &\ge \alpha \mathcal{Q} \left( \sqrt{\frac{\beta MN}{\sigma_c^2 \operatorname{tr}\left[\left(\mathbf{W}_\mathrm{DD}^{{H}} \mathbf{H}_{\mathrm{c,DD}}^{{H}} \mathbf{H}_{\mathrm{c,DD}} \mathbf{W}_\mathrm{DD}\right)^{-1}\right]}}\right)\\ &\triangleq P_e^\mathrm{lb}, \label{lb}
\end{align}}
\hspace{-0.25cm}in which the equality in (\ref{lb}) can be met if and only if $\left[\left(\mathbf{W}_\mathrm{DD}^{{H}} \mathbf{H}_{\mathrm{c,DD}}^{{H}} \mathbf{H}_{\mathrm{c,DD}}\mathbf{W}_\mathrm{DD}\right)^{-1}\right]_{i,i}$, $\forall i \in [1,MN]$ are identical. To facilitate our study, we now adopt $P_e^\mathrm{lb}$ as the new objective function. Furthermore, recall that $\mathcal{Q}(\cdot)$ is monotonically decreasing w.r.t. its argument, which implies that minimizing $P_e^\mathrm{lb}$ is  equivalent to minimizing $\operatorname{tr}\left[\left(\mathbf{W}_\mathrm{DD}^{{H}} \mathbf{H}_{\mathrm{c,DD}}^{{H}} \mathbf{H}_{\mathrm{c,DD}} \mathbf{W}_\mathrm{DD}\right)^{-1}\right]$. Hence,  problem (\ref{pro}) can be approximated as 
\begin{subequations} \label{pro1}
   \begin{align}  
 \min_{\mathbf{W}_\mathrm{DD}} \quad &\operatorname{tr}\left[\left(\mathbf{W}_\mathrm{DD}^{{H}} \mathbf{H}_{\mathrm{c,DD}}^{{H}} \mathbf{H}_{\mathrm{c,DD}} \mathbf{W}_\mathrm{DD}\right)^{-1}\right] \label{obj1}\\
\operatorname{s.t.} \quad & \operatorname{tr}\left(\dot{\mathbf{H}}_\mathrm{s,DD} \mathbf{W}_\mathrm{DD}\mathbf{W}_\mathrm{DD}^{H}\dot{\mathbf{H}}_\mathrm{s,DD}^{{H}}\right) \ge \gamma_1, \label{pe1}\\
&\left[\left(\mathbf{W}_\mathrm{DD}^{{H}} \mathbf{H}_{\mathrm{c,DD}}^{{H}} \mathbf{H}_{\mathrm{c,DD}} \mathbf{W}_\mathrm{DD}\right)^{-1}\right]_{i,i}\le \frac{\beta}{3\sigma_c^2}, \label{II}\\
&\operatorname{tr} (\mathbf{W}_\mathrm{DD}\mathbf{W}_\mathrm{DD}^H)\le P_T, \label{PT1}
\end{align}
\end{subequations}
where $\gamma_1=\sigma_s^2/\gamma_c$ and  (\ref{II}) is employed to guarantee the convexity of $P_e$ and consequently, to establish the validity of $P_e^\mathrm{lb}$. \revise{ We observe that condition (\ref{II}) implies that the approximated problem (\ref{pro1}) is applicable primarily in moderate-to-high SNR scenarios. In lower SNR regimes, the utilization of Jensen's inequality may not hold tightly, potentially leading to less accurate or even invalid solutions, and thus the corresponding precoder design necessitates further consideration, e.g.,  coding techniques \cite{9659787}. }  However, the non-convexity of the newly introduced constraint (\ref{II}) greatly hinders the development of solutions to problem (\ref{pro1}) directly. In what follows, we decouple problem (\ref{pro1}) into two independent sub-problems and then solve them separately.
% According to (\ref{spCRLB}), if there is one single path for sensing, problem (\ref{pro1}) can be simplified as
% \begin{subequations}  \label{prob}
%    \begin{align} 
%  \min_{\mathbf{W}_\mathrm{DD}} \quad &  \left(\frac{1}{\sigma_{{s}}^2} \operatorname{tr}\left(\dot{\mathbf{H}}_\mathrm{s,DD} \mathbf{W}_\mathrm{DD}\mathbf{W}_\mathrm{DD}^{H}\dot{\mathbf{H}}_\mathrm{s,DD}^{{H}}\right)\right)^{-1}\label{obj2}\\
% \operatorname{s.t.} \quad & \text{ (\ref{pe1}), (\ref{II}), (\ref{PT1}}).
% \end{align}
% \end{subequations}
% We observe that

\subsection{Feasibility Checking} \label{iv-b}
Since it is challenging to obtain $\mathbf{W}_\mathrm{DD}$ via solving problem (\ref{pro1}) directly, we adopt the SVD method to  $\mathbf{W}_\mathrm{DD}$ as an alternative, i.e., $\mathbf{W}_\mathrm{DD} =\mathbf{U \Sigma} \mathbf{V}$, 
% \begin{align}
%     \mathbf{W}_\mathrm{DD} =\mathbf{U \Sigma} \mathbf{V}, \label{HIGH}
% \end{align}
where $\mathbf{U}\in \mathbb{C}^{MN\times MN} $ and $\mathbf{V} \in \mathbb{C}^{MN\times MN}$ are both unitary matrices and $\mathbf{\Sigma} \in \mathbb{C}^{MN\times MN}$ is the diagonal singular value matrix. Then, problem (\ref{pro1}) follows that
    \begin{subequations}  \label{prob1}
   \begin{align} 
 \min_{\mathbf{U,\Sigma,V}} \quad & \operatorname{tr}\left[\left(\mathbf{\Sigma}^2 \mathbf{Z}_\mathrm{c}(\mathbf{U})\right)^{-1}\right]\label{obj3} \\
\operatorname{s.t.} \quad & \operatorname{tr}\left(\mathbf{\Sigma}^2 \mathbf{Z}_\mathrm{s}(\mathbf{U})\right)  \ge \gamma_1, \label{pe3}\\
&\left[\mathbf{V}^H\mathbf{\Sigma}^{-1}\mathbf{Z}_\mathrm{c}(\mathbf{U})^{-1}\mathbf{\Sigma}^{-1}\mathbf{V}\right]_{i,i}\le \frac{\beta}{3\sigma_c^2}, \label{each} \\
&\operatorname{tr} (\mathbf{\Sigma^2})\le P_T, \label{power}
\end{align}
\end{subequations}
where $\mathbf{Z}_\mathrm{c}(\mathbf{U})=\mathbf{U}^H {\mathbf{H}}_\mathrm{c,DD}^H  {\mathbf{H}}_\mathrm{c,DD}\mathbf{U}$ and $\mathbf{Z}_\mathrm{s}(\mathbf{U})= \mathbf{U}^H\dot{\mathbf{H}}_\mathrm{s,DD}^H\dot{\mathbf{H}}_\mathrm{s,DD}\mathbf{U}$, respectively. To facilitate our discussion, we represent $\mathbf{W}_\mathrm{s} \mathbf{\Lambda}_\mathrm{s} \mathbf{W}_\mathrm{s}^H$ and $\mathbf{W}_\mathrm{c} \mathbf{\Lambda}_\mathrm{c} \mathbf{W}_\mathrm{c}^H$ as the eigenvalue decomposition of $\dot{\mathbf{H}}_\mathrm{s,DD}^H\dot{\mathbf{H}}_\mathrm{s,DD}$ and ${\mathbf{H}}_\mathrm{c,DD}^H{\mathbf{H}}_\mathrm{c,DD}$,  where $\mathbf{W}_\mathrm{s}$ and $\mathbf{W}_\mathrm{c}$ are regarded as the sensing and communication subspace, respectively \cite{10X}.
% Problem (\ref{prob1}) is still awkward to solve due to highly coupled variables $\mathbf{\Sigma}$ and $\mathbf{U}$ and the existence of (\ref{each}). It is evident that $\mathbf{V}$ only appears in (\ref{each}) and has no impact on both BER and CRLB. Hence,  we can first solve Problem (\ref{prob1}) w.r.t $\mathbf{U}$ and $\mathbf{\Sigma}$ without constraint (\ref{each}) and then find the optimal $\mathbf{U}$. 
Problem (\ref{prob1}) is still hard to solve due to the coupling between the variables $\mathbf{\Sigma}$ and $\mathbf{U}$, as well as the presence of constraint (\ref{each}). However, it becomes evident that the variable $\mathbf{V}$ solely appears in (\ref{each}) and has no discernible impact on both BER and CRB. Consequently, a two-step approach can be proposed to initially determine an optimal $\mathbf{V}$, denoted by $\mathbf{V}^*$, concerning arbitrary $\mathbf{U}$ and $\mathbf{\Sigma}$. Then, one could solve problem (\ref{prob1}) excluding constraint (\ref{each}) for optimizing $\mathbf{U}$ and $\mathbf{\Sigma}$. For notational convenience, the optimal $\mathbf{U}$ and $\mathbf{\Sigma}$ of problem (\ref{prob1}) without constraint (\ref{each}) are denoted as $\mathbf{U}^*$ and $\mathbf{\Sigma}^*$, respectively.

To start with, we address constraint (\ref{each}). Upon a closer inspection, given arbitrary $\mathbf{U}$ and $\mathbf{\Sigma}$, minimizing the largest diagonal element in $\mathbf{V}^H\left(\mathbf{\Sigma } \mathbf{Z}_\mathrm{c}(\mathbf{U}) \mathbf{\Sigma } \right)^{-1}\mathbf{V}$ always produce the best $\mathbf{V}$, despite of the feasibility of (\ref{each}). Thus, we have the following optimization problem
\begin{align}
\nonumber \min_{\mathbf{V}} \ & \max_i \left[\mathbf{V}^H\left(\mathbf{\Sigma } \mathbf{Z}_\mathrm{c}(\mathbf{U}) \mathbf{\Sigma }\right)^{-1}\mathbf{V} \right]_{i,i}, \
\forall i \in [1,MN] \\
&\mathrm{ s.t.} \quad \mathbf{V}^H\mathbf{V}= \mathbf{I}_{MN}.
\label{minmaxc}   
\end{align}
% \begin{align*}
% \nonumber \min_{\mathbf{V}} \ & \max_m \left[\mathbf{V}^H\left(\kappa\sigma_c^2 \mathbf{I}_{MN}+|h_c|^2\mathbf{\Sigma }^2\right)^{-1}\mathbf{V} \right]_{m,m}, \
% \forall m \in [1,MN] \\
% &\mathrm{ s.t.} \quad \mathbf{V}^H\mathbf{V}= \mathbf{I}_{MN}.
% \label{minmaxc}   
% \end{align*} 
% $\frac{\mathrm{tr}\left(\kappa\sigma_c^2 \mathbf{I}_{MN}+|h_c|^2\mathbf{\Sigma }^2\right)^{-1}}{MN}$
To this end, we have the lemma as follows:

\textit{\textbf{Lemma 1:}} 
The minimum objective value of problem (\ref{minmaxc}) is $\frac{\operatorname{tr}(\mathbf{\Sigma Z}_\mathrm{c}(\mathbf{U}) \mathbf{ \Sigma})^{-1}}{MN}$. Furthermore, a qualified candidate of $\mathbf{V}$, which forces  (\ref{minmaxc}) to achieve the minimum value, can be chosen as
\begin{equation}
    \mathbf{V^*}=\boldsymbol{\Upsilon}\mathbf{F}_{MN}, \label{V8}
\end{equation}
in which $\boldsymbol{\Upsilon}$  can be obtained by adopting eigenvalue decomposition  to $ \big(\mathbf{\Sigma Z}_\mathrm{c}(\mathbf{U}) \mathbf{\Sigma}\big)^{-1}$, i.e., $ \big(\mathbf{\Sigma Z}_\mathrm{c}(\mathbf{U}) \mathbf{\Sigma}\big)^{-1}=\boldsymbol{\Upsilon\Psi\Upsilon}^H$, where $\boldsymbol{\Psi}$ is the diagonal eigenvalues matrix. 

\textit{\textbf{Proof:}} Please refer to \cite{ding2003minimum}.

\textbf{\textit{Remark 1:}} In fact, by employing the DFT matrix, such a $\mathbf{V}$ given by (\ref{V8}) forces $\mathbf{V}^H\left(\mathbf{\Sigma Z}_\mathrm{c}\mathbf{(U) \Sigma }\right)^{-1}\mathbf{V}$ to be a circulant matrix with identical diagonal elements. Therefore, the SNR of each received symbol is balanced, such that the lower bound $P_e^\mathrm{lb}$ is achieved in feasible SNR regimes.   We herein emphasize that the choice of $\mathbf{V}$ is generally not unique, for instance, using the inverse normalized DFT matrix $\mathbf{F}_{MN}^H$ to replace $\mathbf{F}_{MN}$ in (\ref{V8}) is also a valid alternative. Moreover, in cases where $MN$ is the power of two, one also has the option of adopting a normalized Hadamard matrix.

In light of Lemma 1 and (\ref{each}), for any given $\mathbf{U}$ and $\mathbf{\Sigma}$, we are capable of concluding that there exists an optimal $\mathbf{V}^*$ given by (\ref{V8}) if the following condition holds 
\begin{equation}
\operatorname{tr}\left(\left( \mathbf{\Sigma Z}_\mathrm{c}\mathbf{(U)\Sigma}\right)^{-1}\right) \le \frac{MN\beta}{3\sigma_c^2}. \label{2MNb}
\end{equation}
\revise{It is observed that $\frac{\operatorname{tr}\left(\boldsymbol{\Sigma} \mathbf{Z}_{\mathrm{c}}(\mathbf{U}) \boldsymbol{\Sigma}\right)^{-1}}{M N}$ is the minimum value of the left-hand side (LHS) of (\ref{each}), which implies that if (\ref{each}) is satisfied, (\ref{2MNb}) must hold well, i.e., (\ref{each}) $\Rightarrow$(\ref{2MNb}). Furthermore, based on Lemma 1, it is always possible to construct an optimal $\mathbf{V}^*$ such that the matrix $\mathbf{V}^{*,H}\mathbf{\Sigma}^{-1}\mathbf{Z}_\mathrm{c}(\mathbf{U})^{-1}\mathbf{\Sigma}^{-1}\mathbf{V}^*$ shares an identical diagonal element $\frac{\operatorname{tr}(\boldsymbol{\Sigma}\mathbf{Z}_\mathrm{c}(\mathbf{U})\boldsymbol{\Sigma})^{-1}}{MN}$. Hence, if (\ref{2MNb}) is fulfilled, it indicates that (\ref{each}) is also capable of being satisfied with the aid of $\mathbf{V}^*$, despite the potential existence of other feasible $\mathbf{V}$ satisfying (\ref{each}), i.e., (\ref{2MNb})$\stackrel{\mathbf{V^*}}{\Rightarrow}$(\ref{each}). Thus, one can employ (\ref{2MNb}) to replace constraint (\ref{each}) in problem (\ref{prob1}) only if it always constructs the optimal $\mathbf{V}^*$ following Lemma 1, such that we now only need to determine $\mathbf{U}$ and $\mathbf{\Sigma}$.}
Furthermore, it can be observed that the LHS of (\ref{2MNb}) is exactly the same as the objective function. In other words, (\ref{2MNb}) implies that the optimal objective value in (\ref{obj3}) should fall below $\frac{MN\beta}{3\sigma_c^2}$, otherwise, constructing a feasible $\mathbf{V}$ is not possible, i.e., (\ref{prob1}) is infeasible.

So far, we have analyzed the feasibility of problem (\ref{prob1}) and the construction of $\mathbf{V}$. All that remains is to obtain 
the solutions $\mathbf{U}^*$ and $\mathbf{\Sigma}^*$ and to check whether 
$ \operatorname{tr}\left(\left( \mathbf{\Sigma^*Z}_\mathrm{c}\mathbf{(U^*)\Sigma^*}\right)^{-1}\right) \le \frac{MN\beta}{3\sigma_c^2} 
$ is satisfied.
If satisfied, there is an optimal $\mathbf{V}^*$ given by (\ref{V8}), which enables the matrix $(\mathbf{V^*})^H\left(\mathbf{\Sigma^* Z}_\mathrm{c}\mathbf{(U^*) \Sigma^* }\right)^{-1}\mathbf{V}^* $ to share an identical diagonal element, i.e., $\frac{\operatorname{tr}(\mathbf{\Sigma^* Z}_\mathrm{c}\mathbf{(U^*) \Sigma^*})^{-1}}{MN}$, ensuring the attainability of the BER lower bound $P_e^{\mathrm{lb}}$, as discussed in (\ref{lb}). 
Given that (\ref{2MNb}) serves as a restriction to the value of the objective function, treating it as a constraint becomes redundant. Thus, we can focus on optimizing matrices $\mathbf{U}$ and $\mathbf{\Sigma}$ by addressing problem (\ref{prob1}), disregarding  (\ref{each}), i.e.,
   \begin{subequations}  \label{svdpro}
   \begin{align} 
 \min_{\mathbf{U,\Sigma}} \quad & \operatorname{tr}\left[\left(\mathbf{\Sigma}^2 \mathbf{Z}_\mathrm{c}(\mathbf{U})\right)^{-1}\right]\label{obj4} \\
\operatorname{s.t.} \quad & \operatorname{tr}\left(\mathbf{\Sigma}^2 \mathbf{Z}_\mathrm{s}(\mathbf{U})\right)  \ge \gamma_1, \ \operatorname{tr} (\mathbf{\Sigma}^2)\le P_T.
\end{align}
\end{subequations}
% where $\mathbf{W_s\Lambda_s W_s}^H$ and $\mathbf{W_c\Lambda_c W_c}^H$ represent the eigenvalue decomposition of $\dot{\mathbf{H}}_\mathrm{s,DD}^H\dot{\mathbf{H}}_\mathrm{s,DD}$ and ${\mathbf{H}}_\mathrm{c,DD}^H{\mathbf{H}}_\mathrm{c,DD}$, respectively. 
 We note that problem (\ref{svdpro}) is feasible if and only if all the contraints are satisfied simultaneously. Thus, given the power budget $ P_T$, it is essential to verify the parameter feasibility of $\gamma_1$ before addressing the problem. Intuitively, due to the power allocation trade-off between sensing and communication, the sensing performance reaches its nadir when the entire power is directed towards downlink transmission, corresponding to BER-only criteria. Conversely, the optimal sensing performance is achieved when allocating all power for tracking the target, which is referred to as CRB-only criteria.  We denote $\gamma_{\max}$ and $ \gamma_{\min}$ as the values of $\gamma_1$ corresponding to the CRB-only criteria and BER-only criteria, respectively. As for the CRB-only criteria, we need to optimize the CRB while constrained by the power budget, i.e.,
\begin{align}    
\nonumber \max_{\mathbf{W}_\mathrm{DD}} \quad &\operatorname{tr}\left(\dot{\mathbf{H}}_\mathrm{s,DD} \mathbf{W}_\mathrm{DD}\mathbf{W}_\mathrm{DD}^{H}\dot{\mathbf{H}}_\mathrm{s,DD}^{{H}}\right)  \\
\operatorname{s.t.} \quad 
&\operatorname{tr} (\mathbf{W}_\mathrm{DD}\mathbf{W}_\mathrm{DD}^H)\le P_T. \label{copr} 
\end{align}
We have the following theorem for revealing the structure of solution to (\ref{copr}):

\textbf{\textit{Theorem 1:}} Problem (\ref{copr}) is convex. The optimal solutions to (\ref{copr}) are $\mathbf{U}_\mathrm{s}^*=\left[\mathbf{w}_{\mathrm{s},1}, \mathbf{0},\dots, \mathbf{0} \right]$ with $\mathbf{w}_{\mathrm{s},1}$ being the eigen vector corresponding to the maximum eigenvalue of $\mathbf{\Lambda}_\mathrm{s}$, $\mathbf{\Sigma}_\mathrm{s}^*=\operatorname{diag}\left( \sqrt{P_T},0,\cdots,0 \right)$, and $\mathbf{V}_\mathrm{s}^*$ is an arbitrary unitary matrix.

 \textbf{\textit{Proof:}} Denote $\mathbf{W}_\mathrm{s,DD}=\mathbf{U}_\mathrm{s}\mathbf{\Sigma}_\mathrm{s}\mathbf{V}_\mathrm{s}$ as the  precoder for CRB-only criteria. Then, problem (\ref{copr}) can be rewritten as 
   \begin{align} 
\max_{\mathbf{U}_\mathrm{s},\mathbf{\Sigma}_\mathrm{s},\mathbf{V}_\mathrm{s}} \quad  \operatorname{tr}\left(\mathbf{\Gamma}_\mathrm{s}\mathbf{W}_\mathrm{s} \mathbf{\Lambda}_\mathrm{s} \mathbf{W}_\mathrm{s}^H\right) \
\operatorname{s.t.} \
\operatorname{tr} (\mathbf{\Sigma}^2_\mathrm{s})\le P_T, \label{power11}
\end{align}
where $\mathbf{\Gamma}_\mathrm{s}=\mathbf{U}_\mathrm{s}\mathbf{\Sigma}_\mathrm{s}^2 \mathbf{U}_\mathrm{s}^H$, whose eigenvalues are represented by $\Gamma_i, i=1,2,\cdots, MN$. The eigenvalues are arranged in a descending order, i.e., $\Gamma_1 \geq \Gamma_2 \geq \cdots \geq \Gamma_{MN}$. Furthermore, the eigenvalues in $\mathbf{\Lambda}_\mathrm{s}$ are denoted as $\Xi_i, i=1,2,\cdots, MN$, which are also arranged in descending order having $\Xi_1 \geq \Xi_2 \geq \cdots \geq \Xi_{MN}$. Note that $\mathbf{\Gamma}_\mathrm{s}$ and $\mathbf{W}_\mathrm{s} \mathbf{\Lambda}_\mathrm{s} \mathbf{W}_\mathrm{s}^H$ are both Hermitian and positive semi-definite.  According to \cite{marshall1979inequalities}, we have the following trace inequality
 \begin{align} 
\left|\operatorname{tr}\left(\mathbf{\Gamma}_\mathrm{s} \mathbf{W}_\mathrm{s} \mathbf{\Lambda}_\mathrm{s} \mathbf{W}_\mathrm{s}^H\right) \right| \overset{(a)}{\leq} \sum_{i=1}^{MN} \Gamma_i \Xi_i \overset{(b)}{\leq} \sum_{i=1}^{MN} \Gamma_i \Xi_1, \label{63b}
\end{align}
where $(a)$ holds if and only if $\mathbf{\Gamma}_\mathrm{s}$ and $\mathbf{W}_\mathrm{s} \mathbf{\Lambda}_\mathrm{s} \mathbf{W}_\mathrm{s}^H$ share the same eigenvectors and $(b)$ holds since $\Xi_1$ is the maximum eigenvalue. It can be verified that the power budget should always be exhausted \cite{liu2021cramer}, which indicates that (\ref{63b}) is upper-bounded by $P_T\Xi_1$. Hence, the maximum objective value for problem (\ref{power11}) is  $P_T\Xi_1$ when $\mathbf{\Gamma}_\mathrm{s}$ and $\mathbf{W}_\mathrm{s} \mathbf{\Lambda}_\mathrm{s} \mathbf{W}_\mathrm{s}^H$ have the same eigenvectors. Then, we have the optimal solutions
\begin{align}
\mathbf{U}_\mathrm{s}^*=[\mathbf{w}_{\mathrm{s},1}, \underbrace {\mathbf{0},\cdots,\mathbf{0}}_{MN-1}] , \ \text{and}  \ \mathbf{\Sigma}_\mathrm{s}^*=\operatorname{diag}(\sqrt{P_T},\underbrace{0, \cdots,0}_{MN-1}), \label{crbonly}
\end{align}
in which $\mathbf{w}_{\mathrm{s},1}$ is the eigenvectors associated with $\Xi_1$. Since $\mathbf{V}_\mathrm{s}$ enters neither the objective function nor the power constraint in (\ref{power11}), we can choose any unitary matrix to be $\mathbf{V}_\mathrm{s}^*$. That completes the proof.

Thus, we have $\gamma_{\max} =\operatorname{tr}\left(\dot{\mathbf{H}}_\mathrm{s,DD} \mathbf{W}_\mathrm{s,DD}^*\mathbf{W}_\mathrm{s,DD}^{*H}\dot{\mathbf{H}}_\mathrm{s,DD}^{{H}}\right)$. On the other hand, to guarantee the CRB constraint is active,  it is required that $\gamma_1$ should be no lower than $\gamma_{\min}$ achieved by the BER-only scheme. In particular, the BER-only scheme is required to solve the following problem
   \begin{align}
        \min_{\mathbf{W}_\mathrm{DD}} 
        \quad  \nonumber &\operatorname{tr}\left[\left(\mathbf{W}_\mathrm{DD}^{{H}} \mathbf{H}_{\mathrm{c,DD}}^{{H}} \mathbf{H}_{\mathrm{c,DD}} \mathbf{W}_\mathrm{DD}\right)^{-1}\right] \\
       \ \operatorname{s.t.} \ \ &\operatorname{tr} (\mathbf{W}_\mathrm{DD}\mathbf{W}_\mathrm{DD}^H)\le P_T. \label{pp2}
    \end{align}
 Based on \cite{9775701}, in the case of transmitting $MN$ symbols within one OTFS block, the optimal minimum BER precoder for the BER-only scheme in the DD domain can be written as 
\begin{align}
    \mathbf{W}_\mathrm{c,DD}^*=\mathbf{W}_\mathrm{c}\sqrt{ \frac{P_T}{\operatorname{tr}\left( \mathbf{\Lambda}_\mathrm{c}^{-1/2} \right)}}\mathbf{\Lambda}_\mathrm{c}^{-1/4}\mathbf{F}_{MN}, \label{bo}
\end{align}
where $\mathbf{\Lambda}_\mathrm{c}^{-1}$ is the inverse of $\mathbf{\Lambda}_\mathrm{c}$ and arranged in descending order. Then, $\gamma_{\min}$ is given by $\operatorname{tr}\left(\dot{\mathbf{H}}_\mathrm{s,DD} \mathbf{W}_\mathrm{c,DD}^*\right.$ $\left.\mathbf{W}_\mathrm{c,DD}^{*H}\dot{\mathbf{H}}_\mathrm{s,DD}^{{H}}\right)$. As a consequence, $\gamma_1$ should satisfy $\gamma_1 \in [\gamma_{\min},\gamma_{\max}]$. From (\ref{bo}), it can be observed that such a precoder prefers to allocate more power to the eigen sub-channels with lower gains. By doing so,  $\mathbf{W}_\mathrm{c,DD}$ endeavors to banlance the SNRs across all sub-channels, and hence, the lower bound $P_e^\mathrm{lb}$ is achievable, which can be characterized as follows 
\begin{align}
P_{e,\min}^\mathrm{lb}=\alpha \mathcal{Q}\left(\sqrt{\frac{\beta MNP_T}{\sigma_c^2\left(\operatorname{tr}\left( \mathbf{\Lambda}_\mathrm{c}^{-1/2}\right)\right)^2}} \right) \label{pem}.
\end{align}
 Based on the above discussions, we observe that $\mathbf{W}_\mathrm{s,DD}^*$ is a rank-$1$ matrix while $\mathbf{W}_\mathrm{c,DD}^*$ is with full rank, which indicates $\mathbf{W}_\mathrm{s,DD}^*$ will allocate the entire power to transmit one symbol in each OTFS block. This is different from the BER-only criteria due to the fact that $\mathbf{W}_\mathrm{s,DD}^*$ does not care how much information is reliably conveyed. By doing so, the transmitted symbol has the maximized SNR, resulting in an enhanced sensing performance. Observing (\ref{pem}), it can be verified that with the decrease of transmitted symbol numbers, the average received SNR at the UE is improved, resulting in a better BER performance \cite{9775701}.  Furthermore, by comparing (\ref{crbonly})-(\ref{pem}), we can observe the following corollary.

\textbf{\textit{Corollary 1:}} Both the BER and CRB achieve their corresponding optimum when an OTFS block only consists of a single symbol to be transmitted. Furthermore, the optimal precoder at this point is 
\begin{equation}
\mathbf{w}_\mathrm{dd}=\left\{\begin{array}{l}
\sqrt{P_T} \mathbf{w}_\mathrm{c,1}, \text { if } P_T\left|\mathbf{w}_\mathrm{c,1}^H \mathbf{w}_\mathrm{s,1}\right|^2> \gamma_1/\Xi_1, \\
x\mathbf{w}_\mathrm{s,1}+y \mathbf{w}_\mathrm{u}, \text { otherwise, }
\end{array}\right.
\end{equation}
with 
\begin{align}
    x=\sqrt{\frac{\gamma_1}{|\Xi_1|}} \frac{\mathbf{w}_{\mathrm{s},1}^H\mathbf{w}_{\mathrm{c},1}}{|\mathbf{w}_{\mathrm{s},1}^H\mathbf{w}_{\mathrm{c},1}|}, \ \text{and} \ y=\sqrt{P_T-\frac{\gamma_1}{|\Xi_1|}} \frac{\mathbf{w}_{\mathrm{u}}^H\mathbf{w}_{\mathrm{c},1}}{|\mathbf{w}_{\mathrm{u}}^H\mathbf{w}_{\mathrm{c},1}|}, 
\end{align}
where $\mathbf{w}_\mathrm{c,1}$ is the eigenvector associated with the largest eigenvalue in $\mathbf{\Lambda}_\mathrm{c}$, and $\mathbf{w}_{\mathrm{u}} =\frac{\mathbf{w}_{\mathrm{c},1}-(\mathbf{w}_{\mathrm{s},1}^H\mathbf{w}_{\mathrm{c},1})\mathbf{w}_{\mathrm{s},1}}{||\mathbf{w}_{\mathrm{c},1}-(\mathbf{w}_{\mathrm{s},1}^H\mathbf{w}_{\mathrm{c},1})\mathbf{w}_{\mathrm{s},1}||}$.
% is obtained by Gram-Schmidt orthogonalization.  

 \textbf{\textit{Proof:}} Please refer to Appendix \ref{APPA}

\textbf{\textit{Remark 2:}} 
% In addition to the maximization of symbol SNR achieved by transmitting a single symbol, Theorem 2 may also be construed that the number of transmitted symbols affects the inherent randomness. This reduction in randomness, in turn, enhances the symbol detection accuracy at the receiver and the overall sensing performance.
It is noteworthy that although corollary 1 indicates that both the BER and CRB have the same tendency regarding the number of transmitted symbols, it does not contradict the trade-off between communication and sensing. Once the number of symbols for an OTFS block is fixed, the non-trivial trade-off is unavoidable. This is because the dual-functional precoder not only needs to allocate power for balancing the SNRs across the selected communication eigen sub-channels but also entails the power concentration to transmit one symbol for improving the sensing performance. We will discuss the case of transmitting $MN$ symbols in Sec. \ref{scde}. Besides, the consistency of this tendency comes at the expense of reduced spectral efficiency.
% Note that the eigenvalues in $\mathbf{\Lambda_s}$ and $\mathbf{\Lambda_c}$ are arranged in descending order.
\subsection{Low Complexity DD Precoder Design } \label{scde}
In this subsection, we propose a low computational complexity method to devise the DD precoder by solving problem (\ref{svdpro}). We first consider a simple scenario where the signal propagation between the transmitter and UE is also assumed to be dominated by the LoS link, indicating $\mathbf{H}_{\mathrm{c,DD}}^{{H}} \mathbf{H}_{\mathrm{c,DD}}=|h_c|^2\mathbf{I}_{MN}$ with $h_c$ being the complex path gain. Consequently, problem (\ref{svdpro}) becomes
\begin{subequations} 
    \begin{align}
 \ \min _{\mathbf{U}, \mathbf{\Gamma}} & \ \operatorname{tr}\left(\frac{1}{|h_c|^2}\mathbf{\Gamma}^{-1}\right) \label{trber}\\
\mathrm{ s.t. } & \ \operatorname{tr}\left(\boldsymbol{\Gamma}\right) \le P_T, \ \text{and} \ \operatorname{tr}\left( \mathbf{\Gamma}\mathbf{Z}_\mathrm{s}\left(\mathbf{U}\right)\right)\ge {\gamma _1}, 
  \end{align} \label{scp}
\end{subequations}
\hspace{-1.3mm}in which $\mathbf{\Gamma}=\mathbf{\Sigma}^2$. Although problem (\ref{scp}) is non-convex, it can be addressed by first solving the optimal $\mathbf{\Gamma}$ given a $\mathbf{U}$, followed by determining of the optimal $\mathbf{U}$. For a given $\mathbf{U}$, the Lagrangian function of (\ref{scp}) is expressed as
% and can be solved by CVX via the interior-point method \cite{grant2014cvx}. However, the interior-point method is with high computational complexity, i.e., $\mathcal{O}(MN)^{3.5}$. To obtain a low-cost solution, we first derive the Lagrangian function given by
 \begin{align}
   \mathcal{L}(\mathbf{\Gamma},\lambda,\mu)=\mathrm{tr}\left(\left(|h_c|^2\mathbf{\Gamma}\right)^{-1}\right)+\mu(\mathrm{tr}(\mathbf{\Gamma})-P_T) -\lambda \left(\mathrm{tr}(\mathbf{\Gamma}\mathbf{Z}_\mathrm{s}(\mathbf{U})-{\gamma_1}\right), \label{la1g}
\end{align}
in which $\lambda \ge 0$ and $\mu \ge 0$ are the associated Lagrangian multipliers.
By setting the gradient w.r.t. $\mathbf{\Gamma}$ to zero, we obtain the optimal $\mathbf{\Gamma}$ given by
 \begin{align}
\mathbf{\Gamma}^*=\frac{\left(\mu\mathbf{I}_{MN}-\lambda\mathbf{Z}_\mathrm{s}(\mathbf{U})\right)^{-\frac{1}{2}}}{|h_c|}.\label{gamma}
\end{align}
It is found that the optimal $ \mathbf{Z}_\mathrm{s}(\mathbf{U})$ must exhibit a diagonal structure as $\mathbf{\Gamma}^*$ is diagonal. In other words, an optimal choice of $\mathbf{U}$ entails ensuring that $\mathbf{U}^H\mathbf{W}_\mathrm{s}$ forms a permutation matrix. Thus, $\mathbf{U}$ can be chosen as $\mathbf{W}_\mathrm{s}$\footnote{ As per (\ref{gamma}), it is worth noting that the optimal $\mathbf{U}$ may not be uniquely determined in this scenario. For instance, alternative choices for $\mathbf{U}$ can be derived by rearranging the column order within $\mathbf{W}_\mathrm{s}$. Given that $\mathbf{U}^H\mathbf{W}_\mathrm{s}$ exclusively decides the element order in $\mathbf{\Lambda}_\mathrm{s}$, the optimal $\mathbf{\Gamma}$ will be designed following the same order as $\mathbf{\Lambda}_\mathrm{s}$ without changing the values of $\mathbf{\Gamma}$, which has no impact on the objective value of (\ref{trber}). However, the optimal $\mathbf{\Gamma}$ is unique for a given $\mathbf{U}$.}. 
Then, we need to solve the values of $\lambda$ and $\mu$. 
Based on the KKT conditions, we can immediately observe that $\mu>0$ since the available power budget should always be exhausted. Furthermore, if the CRB constraint is inactive, i.e., $\lambda=0$, we can obtain the optimal $\mathbf{\Gamma}=(P_T/MN) \mathbf{I}_{MN}$ since all the eigen sub-channels have an identical gain $|h_c|^2$.  However, suppose that $\lambda >0$, which indicates the CRB constraint is active. In this case, it is intractable to obtain $\lambda$ and $\mu$ directly, while efficient iterative methods have been proposed to obtain these parameters, e.g., the gradient descent method. 
% \cite{ruder2016overview}.
For more details, interested readers can refer to our previous work \cite{wu2023optimal}.
% According to (\ref{la1g}), we have the following dual problem 
% \begin{align}
%  \ \max _{\lambda \geq 0, \mu \geq 0} \inf _{\mathbf{\Gamma} \succeq \mathbf{0}} \mathcal{L}\left(\mathbf{\Gamma}, \lambda, \mu\right) \triangleq \max _{\lambda \geq 0, \mu \geq 0} \mathcal{D}\left(\lambda, \mu\right), 
% \end{align}
% where $\mathcal{D}\left(\lambda, \mu\right)$ represents the dual function, which can be obtained by substituting (\ref{gamma}) into (\ref{la1g}). Since $ (\mathcal{P}3)$ is convex given $\mathbf{U}$ and satisfies the Slater's conditions, the dual gap between $ (\mathcal{P}3)$ and $ (\mathcal{P}4)$ is zero. To find the optimal $\lambda$ and $\mu$, one can apply the gradient descent method \cite{5406097}, i.e.,
% \begin{align}
%     \lambda^r=\left[\lambda^{r-1}+\epsilon_1\nabla \mathcal{D}\left(\lambda^{r-1}\right)\right]^{+},\\
%      \mu^r=\left[\mu^{r-1}+\epsilon_2\nabla \mathcal{D}\left(\mu^{r-1}\right)\right]^{+},
% \end{align}
% where $r$ denotes the iteration index, $\epsilon_1$ and $\epsilon_2$ are the corresponding stepsize \cite{boyd2004convex}. 
Relying on the aforementioned discussions, we observe that $\mathbf{W}_\mathrm{DD}$ is only required to align the sensing subspace $\mathbf{W}_\mathrm{s}$, at this stage, since the communication subspace $\mathbf{W}_\mathrm{c}$ is reduced to $\mathbf{I}_{MN}$. 

In a more general scenario involving multi-path-based communication, i.e., when $\mathbf{H}_{\mathrm{c,DD}}^{{H}} \mathbf{H}_{\mathrm{c,DD}}$ is no longer diagonal, it is intractable to determine $\mathbf{U}$ and $\mathbf{\Sigma}$ separately due to the interaction of $\mathbf{W}_\mathrm{c}$ and $\mathbf{W}_\mathrm{s}$.
% Because of the strong coupling of $\mathbf{\Sigma}$  and $\mathbf{U}$, it is difficult to obtain the optimal values by solving problem (\ref{svdpro}) directly. 
To overcome this challenge, we resort to optimizing the covariance matrix $\mathbf{W}_\mathrm{DD}\mathbf{W}_\mathrm{DD}^H=\mathbf{U}\mathbf{\Sigma}^2\mathbf{U}^H$, denoted by $\mathbf{P}$. Thus, we have
\vspace{-0.2cm}
\begin{subequations} \label{svdpro1}
    \begin{align}
        \min_{\mathbf{P}\succeq \mathbf{0}} \quad &   \operatorname{tr}\left[\left(\mathbf{P}\mathbf{W}_\mathrm{c}\mathbf{\Lambda}_\mathrm{c}\mathbf{W}_\mathrm{c}^H\right)^{-1}\right] \\
        \operatorname{s.t.}  \quad &\operatorname{tr}\left(\mathbf{P} \mathbf{W}_\mathrm{s}\mathbf{\Lambda}_\mathrm{s}\mathbf{W}_\mathrm{s}^H\right) \ge \gamma_1,  \operatorname{tr}\left( \mathbf{P}\right) \le P_T,
    \end{align}
\end{subequations}
It is found that problem (\ref{svdpro}) is convex and thus can be solved by standard solvers using the interior-point method, such as CVX \cite{grant2014cvx}. 
\revise{Nevertheless, we note that the Hermitian matrix $\mathbf{P}$ contains $MN$ real variables and $\frac{(MN)^2-MN}{2}$ complex variables, i.e., there are $(MN)^2$ real variables to be determined in total. In each iteration, the interior-point method introduces a significant computational complexity of $\mathcal{O}((MN)^{7})$  \cite{grant2014cvx}. Given the convergence tolerance threshold $\xi_{0}$, the total computational complexity is given by $\mathcal{O}(\log{\frac{1}{\xi_0}}(MN)^{7})$. }
To address (\ref{svdpro}) at low overhead, we then express the Lagrangian function as follows
\begin{align}
 \nonumber  \mathcal{L}\left( \mathbf{P},\lambda,\mu \right)=\operatorname{tr}\left[\left(\mathbf{P} \mathbf{W}_\mathrm{c}\mathbf{\Lambda}_\mathrm{c}\mathbf{W}_\mathrm{c}^H\right)^{-1}\right]+\mu\left(\operatorname{tr}\left( \mathbf{P}\right) - P_T \right) -\lambda\left( \operatorname{tr}\left(\mathbf{P}\mathbf{W}_\mathrm{s}\mathbf{\Lambda}_\mathrm{s}\mathbf{W}_\mathrm{s}^H\right) - \gamma_1 \right). \label{lag}
\end{align}
Then, we have the following KKT conditions
\begin{equation} \label{kkt}
    \begin{array}{rrrr}
         \frac{\partial \mathcal{L}}{\partial \mathbf{P}} &=\mathbf{0}, \\
         \operatorname{tr}\left( \mathbf{P}\right) \le P_T, \ \mu\left(\operatorname{tr}\left( \mathbf{P}\right) - P_T \right)&=0,\\
      \lambda\left( \operatorname{tr}\left(\mathbf{P} \mathbf{W}_\mathrm{s}\mathbf{\Lambda}_\mathrm{s}\mathbf{W}_\mathrm{s}^H\right) - \gamma_1 \right)&=0,
      \\ \operatorname{tr}\left(\mathbf{P} \mathbf{W}_\mathrm{s}\mathbf{\Lambda}_\mathrm{s}\mathbf{W}_\mathrm{s}^H\right) &\ge \gamma_1.
      \end{array}
\end{equation}
By some algebraic manipulations, we have $\frac{\partial \mathcal{L}}{\partial \mathbf{P}}=-\lambda \left( \mathbf{W}_\mathrm{s}\mathbf{\Lambda}_\mathrm{s}\mathbf{W}_\mathrm{s}^H\right)-\left(\mathbf{P} \mathbf{W}_\mathrm{c}\mathbf{\Lambda}_\mathrm{c}\mathbf{W}_\mathrm{c}^H\mathbf{P} \right)^{-1}+\mu \mathbf{I}=\mathbf{0}$, and hence the optimal $\mathbf{P}$ can be straightforwardly derived by
\begin{equation}
    \mathbf{P^*} =\mathbf{S}_0^{-1} \left( \mathbf{S}_0 \mathbf{S}_1 \mathbf{S}_0  \right)^{\frac{1}{2}} \mathbf{S}_0^{-1},
\label{opti}
\end{equation}
  % \left(\frac{\mu}{\lambda}\mathbf{W_c\Lambda_c W_c}^H-\frac{1}{\sigma_{s}^2\lambda} \mathbf{W_c\Lambda_c W_c}^H\mathbf{W_s\Lambda_s W_s}^H \right)^{-\frac{1}{2}}. 
 where $\mathbf{S}_0=\mathbf{W}_\mathrm{c}\mathbf{\Lambda}_\mathrm{c}^{\frac{1}{2}} \mathbf{W}_\mathrm{c}^H$ and $\mathbf{S}_1= \mathbf{W}_\mathrm{s}\left(\mu \mathbf{I}_{MN}-\lambda \mathbf{\Lambda}_\mathrm{s}\right)^{-1} \mathbf{W}_\mathrm{s}^H$. From (\ref{kkt}), $\mu>0$ holds since the allocated power budget is exhausted similar to (\ref{la1g}). Given a suitable $\gamma_1\in [\gamma_{\min},\gamma_{\max}]$, $\lambda$ is also ensured to be positive. Furthermore, $\mathbf{P}$ should be full-rank due to the matrix inverse operation, otherwise (\ref{lag}) is unbounded, which indicates $\left(\mu \mathbf{I}_{MN}-\lambda \mathbf{\Lambda}_\mathrm{s}\right)$ is a positive definite matrix, i.e., $\mu>\lambda\Xi_1$.
% However, it should be noted that if $\lambda=0$, i.e., the BER constraint is inactive, the first term of (\ref{kkt}) loses significance, resulting in the Lagrangian multiplier method being invalid and inapplicable, which we will discuss later.  Furthermore, suppose that $\lambda>0$, which indicates the BER constraint is active, 
As observed, (\ref{opti}) reveals an inherent compromise that $\mathbf{P^*}$ strikes a balance situated at the trade-off between the sensing subspace $\mathbf{W}_\mathrm{s}$ and the communication subspace $\mathbf{W}_\mathrm{c}$, as well as the power allocation \cite{10X}. 
%  In particular, in cases where the signal propagation between the transmitter and UE is also dominated by a LoS path, leading to ${\mathbf{H}}_\mathrm{c,DD}^H{\mathbf{H}}_\mathrm{c,DD}=|h_c|^2 \mathbf{I}_{MN}$ with $h_c$ being the complex LoS path gain, we have the simplified $\mathbf{P}^*$ expressed as
% \begin{equation}
%     \mathbf{P^*} = \mathbf{W_s}\left(\frac{\mu |h_c|^2}{\lambda} \mathbf{I}_{MN}-\frac{|h_c|^2}{\sigma_{s}^2\lambda} \mathbf{\Lambda_s}\right)^{-\frac{1}{2}} \mathbf{W_s}^H, \label{optii}
% \end{equation}
% which indicates that the communication subspace at this point is the entire space and the precoder only requires to align the sensing subspace $\mathbf{W_s}$. 

As a step forward, we need to calculate the values of $\lambda$ and $\mu$. Based on (\ref{lag}), we have the dual problem of (\ref{svdpro1}) as follows
\begin{align}
    \max_{\lambda\ge0,\mu\ge0,\mathbf{P}\succeq\mathbf{0}} \inf \ \mathcal{L}\left({\mathbf{P},\lambda,\mu}\right) \triangleq \max_{\lambda\ge0,\mu\ge0} \mathcal{D}(\lambda,\mu), \label{dula}
\end{align}
where $\mathcal{D}(\lambda,\mu)$ denotes the dual function.  It is evident that problem (\ref{svdpro1}) and the dual problem  (\ref{dula}) have an identical solution, i.e., the dual gap is zero as (\ref{svdpro1}) is convex and satisfies Slater's conditions. To find the optimal $\lambda$ and $\mu$, we can adopt the ellipsoid method. According to \cite{grotschel1981ellipsoid}, an ellipsoid shape $E\left(\boldsymbol{\kappa},\mathbf{B} \right)$ can be represented by
\begin{align}
    E\left( \boldsymbol{\kappa},\mathbf{B}\right) \triangleq \left\{ \boldsymbol{\theta}: (\boldsymbol{\theta-\kappa})^T\mathbf{B}(\boldsymbol{\theta-\kappa}) \le1 \right\},
\end{align}
where $\boldsymbol{\kappa}$ is the ellipsoid center denoted by $\boldsymbol{\kappa}=[\lambda,\mu]^T$ and $\mathbf{B}$ represents a positive semidefinite symmetric matrix. Denote $\mathbf{d}=[d_{\lambda}, d_{\mu}]^T$ as the subgradient of $\mathcal{D}(\lambda,\mu)$ w.r.t the center $\boldsymbol{\kappa}$, which is given by
\begin{align}
     \nonumber d_{\lambda}&=-\operatorname{tr}\left(\mathbf{P^*} \mathbf{W}_\mathrm{s} \mathbf{\Lambda}_\mathrm{s} \mathbf{W}_\mathrm{s}^H\right) + \gamma_1, \\
    d_{\mu}&=\operatorname{tr}(\mathbf{P^*})-P_T.
\end{align}
To execute the ellipsoid method, it is required to initialize an ellipsoid, represented by $E(\boldsymbol{\kappa}_0,\mathbf{B}_0)$ that can involve all feasible candidates of $\lambda$ and $\mu$. Then, one can  mathematically update the ellipsoid method as follows
\begin{align}
\mathbf{\tilde{d}}_r&=\frac{\mathbf{d}_r}{\sqrt{\mathbf{d}_r^T\mathbf{B}_r^{-1}\mathbf{d}_r}}, \label{dr}\\
       \boldsymbol{\kappa}_{r+1}&=\boldsymbol{\kappa}_{r}-\frac{1}{3}\mathbf{B}_r^{-1}\mathbf{\tilde{d}}_{r},\label{kr}\\
       \mathbf{B}_{r+1}^{-1}&= \frac{4}{3}\left( \mathbf{B}_r^{-1} -\frac{2}{3}\mathbf{B}_r^{-1}\mathbf{\tilde{d}}_r\mathbf{\tilde{d}}_r^T\mathbf{B}_r^{-1}\right), \label{br}
\end{align}
in which $r$ is the iteration index. The $(r+1)$-th iteration discards one half of the ellipsoid from the $r$-th iteration, determined by the vector $\mathbf{d}_{r+1}$, and simultaneously introduces a new minimal ellipsoid containing the complementary half, where the optimal values for $\lambda$ and $\mu$ reside. (\ref{dr})-(\ref{br}) are repeated until $\sqrt{\mathbf{d}_r^T\mathbf{A}_r^{-1}\mathbf{d}_r}<\xi_0$ to guarantee the convergence \cite{1658226}. Based on the obtained $\mathbf{P}^*$ at hand, it is readily to derive the optimal $\mathbf{U}^*$ and $\mathbf{\Sigma}^*$ via eigenvalue decomposition. 

Based on the above discussions, the dual functional precoder design in the DD domain can be solved by first optimizing $\mathbf{U}$ and $\mathbf{\Sigma}$ via solving problem (\ref{svdpro1}), in which the parameter $\gamma_1$ is pre-determined by addressing problem (\ref{copr}) and (\ref{pp2}). Then, we need to perform the feasibility checking via (\ref{2MNb}). If the feasibility is satisfied, $\mathbf{V}$ can be constructed through Lemma 1\footnote{\revise{We highlight that the proposed approach guarantees a globally optimal solution to the original problem only when the condition \( \operatorname{tr}\left(\left( \mathbf{\Sigma^*Z}_\mathrm{c}\mathbf{(U^*)\Sigma^*}\right)^{-1}\right) \le \frac{MN\beta}{3\sigma_c^2} \) is satisfied. This condition implies that our proposed approach is effective primarily at high SNR regimes, as detailed in (\ref{high}). Conversely, when this condition is not met, it becomes impossible for our proposed method to obtain the optimal \(\mathbf{V}^* \) based on Lemma 1, such that the approximated problem (\ref{prob1}) is infeasible, leading to the failure to achieve the intended global optimum of the original problem.}}. The overall procedures for devising the precoder in the DD domain are summarized in \textbf{Algorithm 1}. \revise{ We proceed to analyze the computational complexity of the proposed algorithm. Initially, calculating \(\mathbf{P}^*\) via (\ref{opti}) involves matrix inversion and square root operations, resulting in a cubic complexity of \(\mathcal{O}((MN)^3)\). Furthermore, the dual variables \([ \lambda,\mu]^T\) are updated exploiting the ellipsoid method, with each iteration incurring a complexity of \(\mathcal{O}(4)\) \cite{8352726}. Therefore, the overall computational complexity of \textbf{Algorithm 1} is \(\mathcal{O}\big(((MN)^3 + 4)\ln(1/\xi_0)\big)\), which is notably lower than that of the interior-point method.}

\renewcommand{\algorithmicrequire}{\textbf{Initialization}} 
\renewcommand{\algorithmicensure}{\textbf{Output:}} 
\renewcommand{\algorithmicrepeat}{\textbf{Repeat:}}
\renewcommand{\algorithmicuntil}{\textbf{Until:}}
\renewcommand{\thefootnote}{1}
\begin{algorithm}[t]
  \caption{ The Overall Procedure for Solving Problem (\ref{prob1})} % 
  \label{algorithm1}
  \begin{algorithmic}[1]
    \State
      \textbf{Input:} $\mathbf{H}_{\mathrm{c,DD}}$, $\mathbf{\dot{H}}_{\mathrm{s,DD}}$, $\xi_0$, $P_T$, $\gamma_1$, $M$, $N$, $\alpha$, $\beta$, $\sigma_c^2$, $\sigma_s^2$
    \State \textbf{Initialization:} $E\left(\boldsymbol{\kappa}_0,\mathbf{B}^{-1}_0\right)$, $r=0$
    \Repeat  
      \State Given $\boldsymbol{\kappa}_r=[\lambda_r,\mu_r]^T$, calculate $\mathbf{P}^*_r$ and the subgradient $\mathbf{\tilde{d}}_r$ based on (\ref{opti}) and (\ref{dr}), respectively.        
       \State Update the ellipsoid $E\left( \boldsymbol{\kappa}_{r+1},\mathbf{B}_{r+1}^{-1}\right)$ by computing (\ref{kr}) and (\ref{br}), respectively.
       \State Update $r=r+1$.          
    \Until{$\sqrt{\mathbf{d}_r^T\mathbf{A}_r^{-1}\mathbf{d}_r}<\xi_0$}
      \State  Perform the feasibility check via (\ref{2MNb}).
      \If {(\ref{2MNb}) is satisfied,}
       \State $\mathbf{V}^*$ is obtained by (\ref{V8}).
     \EndIf
     \State \textbf{Output:} $\mathbf{W}_\mathrm{DD}^*=\mathbf{U^*\Sigma^*V^*}$
  \end{algorithmic}
\end{algorithm} 

\section{Simulation Results}
\label{sec5}
 In this section, we present the simulation results to verify the effectiveness of our developed method. The system setup is as follows:  we set $M=8$, $N=8$ with $4$-QAM signaling and the carrier frequency is $4$ GHz with a sub-carrier spacing of $2 $ kHz. The maximum speed of the mobile user is $v_{\max }=120 \mathrm{~km} / \mathrm{h}$, leading to a maximum Doppler-shift tap $k_{\max }=2$. The maximum delay tap is assumed to be $l_{\max }=4$, such that the delay and Doppler indices are limited in $[0,l_{\max}]$ and $[-k_{\max},k_{\max}]$ \cite{10138552}.  The noise covariances are set to $\sigma_s^2=\sigma_c^2=0$ dBm. As for communication, we consider a time-variant channel having $P=3$ resolvable paths. We set the fading coefficients are independently complex Gaussian distributed with a zero-mean and a variance of $1/P$. 
 % The required CRB threshold is set to $5\times 10^{-5}$.
 We choose $\xi_0=10^{-3}$ as the stop iteration criterion. 
 % All the results are obtained by adopting Monte Carlo with $500$ runs.

\begin{figure}
  \begin{subfigure}{0.47\linewidth}
\includegraphics[width=\textwidth]{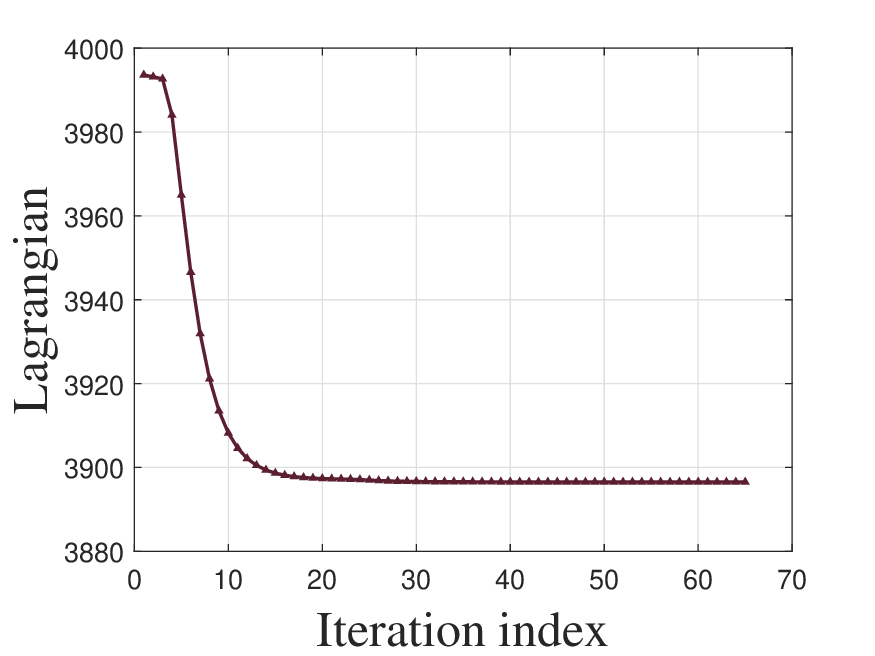}
       \caption{\revise{Lagrangian.}}
    \label{figsub1}
  \end{subfigure}%
  \begin{subfigure}{0.47\linewidth}
\includegraphics[width=\textwidth]{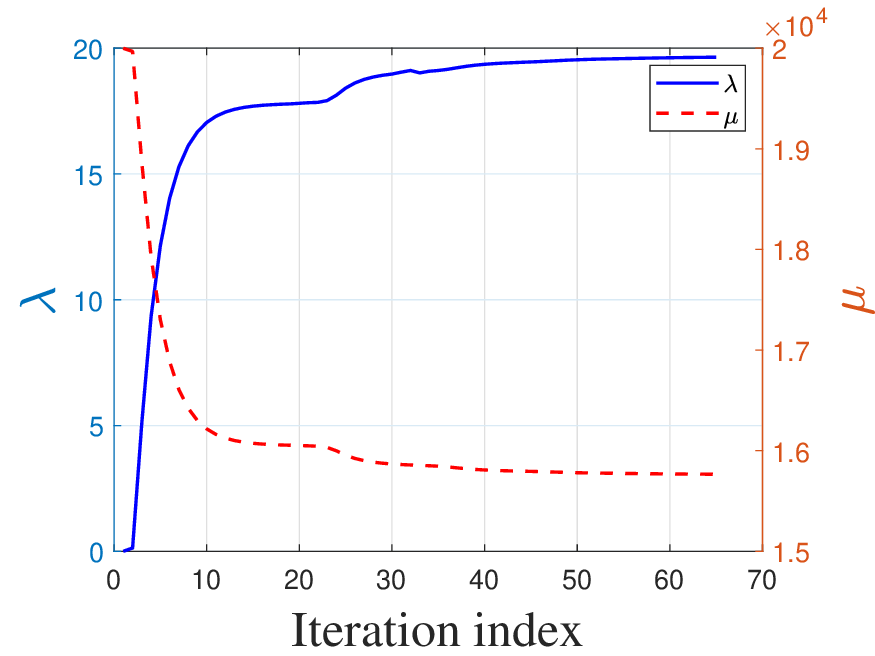}
    \caption{\revise{$\lambda$ and $\mu$.}}
    \label{figsub2}
  \end{subfigure}
  \captionsetup{font=small,name=Fig, labelsep=period,justification=raggedright,singlelinecheck=false}
    \caption[center]{ \revise{Convergence behavior of the proposed \textbf{Algorithm 1} under $P_T= 30$ dBm.}}
  \label{fig2} 
\end{figure}

 Before comparing the system performance,  we first depict the convergence behavior of the proposed low-complexity method under the case of $P_T=30$ dBm.  From Fig. \ref{figsub1}, we observe that the averaged Lagrangian value, as defined in (\ref{lag}), decreases fast with \textbf{Algorithm 1} proceeding.  In Fig. \ref{figsub2}, it is found that the averaged values of multipliers $\lambda$ and $\mu$ change quickly, and finally our proposed algorithm converges in about $60 $ iterations, which guarantees the low-complexity precoder design. \revise{Furthermore, Fig. 4 presents the computation time ratios of the interior point method compared to our proposed method, with the product of \(M\) and \(N\) varying from $16$ to $256$. It is observed that the proposed method significantly reduces computation time compared to the interior point method, especially as the product of \(M\) and \(N\) increases, demonstrating a superior computational efficiency of the proposed approach, particularly in scenarios involving larger values of \(M\) and \(N\). } 

\begin{figure}
    \centering
    \includegraphics[width=0.6\columnwidth]{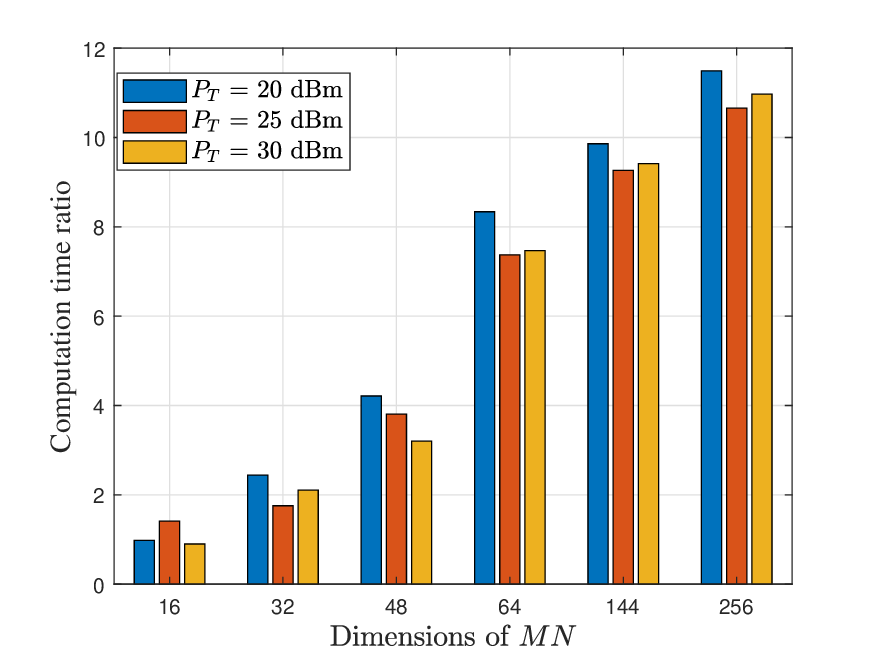}
    \captionsetup{font=small,name=Fig, labelsep=period,justification=raggedright,singlelinecheck=false}
    \caption{ \revise{ Computation time ratios versus the product of $M$ and $N$.}}
    \label{figcompu} 
\end{figure}

In order to evaluate the communication performance of the proposed method, we show the BER performance by varying different power budgets given the CRB threshold. Furthermore, we provide three benchmark schemes to compare with the proposed method for verifying the system performance, i.e.,
\begin{itemize}
\item Benchmark Scheme $1$ (Lower bound): This method is the BER-only scheme, where the CRB constraint (\ref{pe1}) is ignored. The optimal precoder is obtained by solving problem (\ref{pp2}). Also, the BER is available in (\ref{pem}).
    \item Benchmark Scheme $2$ (ZF scheme): This approach adopts a ZF equalizer \cite{10138552} at the receiver without employing a precoder at the transmitter, i.e., $\mathbf{W}_{\mathrm{DD}}=\mathbf{I}_{MN}.$ Also, the CRB constraint is omitted in this method.
    \item Benchmark Scheme $3$ (MMSE scheme): A MMSE equalizer \cite{palomar2005minimum} is adopted at the receiver without taking into account the CRB constraint and the precoder, i.e., $\mathbf{Q}=\left(\sigma_c^2\mathbf{I}_{MN}+\mathbf{W}_\mathrm{DD}^{{H}} \mathbf{H}_{\mathrm{c,DD}}^{{H}} \mathbf{H}_{\mathrm{c,DD}} \mathbf{W}_\mathrm{DD}\right)^{-1}$ $\mathbf{W}_\mathrm{DD}^H\mathbf{H}_{\mathrm{c,DD}}^H$ with $\mathbf{W}_\mathrm{DD}=\mathbf{I}_{MN}$.
\end{itemize}
\begin{figure}
    \centering
\includegraphics[width=0.6\columnwidth]{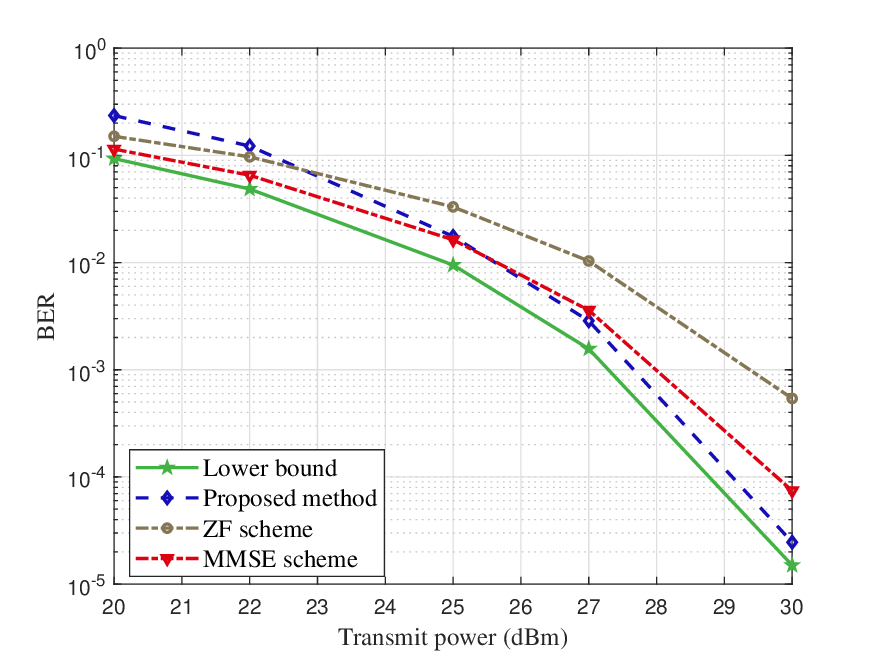}
\captionsetup{font=small,name=Fig, labelsep=period,justification=raggedright,singlelinecheck=false}
    \caption{The BER performance with different transmit power budgets given the required BER threshold $\gamma_c=5\times 10^{-5}$.}
    \label{fig4}
\end{figure}
In Fig. \ref{fig4}, although the MMSE scheme demonstrates superior BER performance compared to the ZF scheme,  it is noteworthy that both two methods marginally impact the BER performance as they solely adopt an equalization for detection, lacking a well-designed precoder. \revise{Additionally, given the required CRB threshold, we observe that the proposed method exhibits a less favorable BER performance compared to the ZF and MMSE approaches in cases of lower transmit power budgets. This is due to the fact that a poor power budget renders the CRB constraint more stringent and thus the optimal precoder has to sacrifice a certain of BER performance for achieving the desired sensing performance. In contrast, the ZF and MMSE schemes, which do not account for the sensing constraints, demonstrate better BER performance in the low SNR regime. It is noted that this issue can be mitigated if the sensing channel $\mathbf{H}_\mathrm{s,DD}$ is strong or if the CRB threshold is relaxed. Furthermore, as the power budgets become generous, our proposed method is capable of surpassing both the ZF and MMSE methods ($P_T\ge 23$ dBm for the ZF scheme and  $P_T\ge 25$ dBm for the MMSE scheme) and becomes closer to the performance lower bound.} These results are as expected since a more generous power budget allows for the fulfillment of the pre-set CRB threshold with primary consideration on the BER performance. Notably, since $\mathbf{W}_{\mathrm{DD}}$ has to allocate a part of power to ensure the CRB threshold is satisfied, this indicates that the Doppler shift estimation accuracy of our proposed method will always be better than that of the ZF and MMSE schemes. 

\begin{figure}
    \centering
\captionsetup{font=small,name=Fig, labelsep=period,justification=raggedright,singlelinecheck=false}\includegraphics[width=0.6\columnwidth]{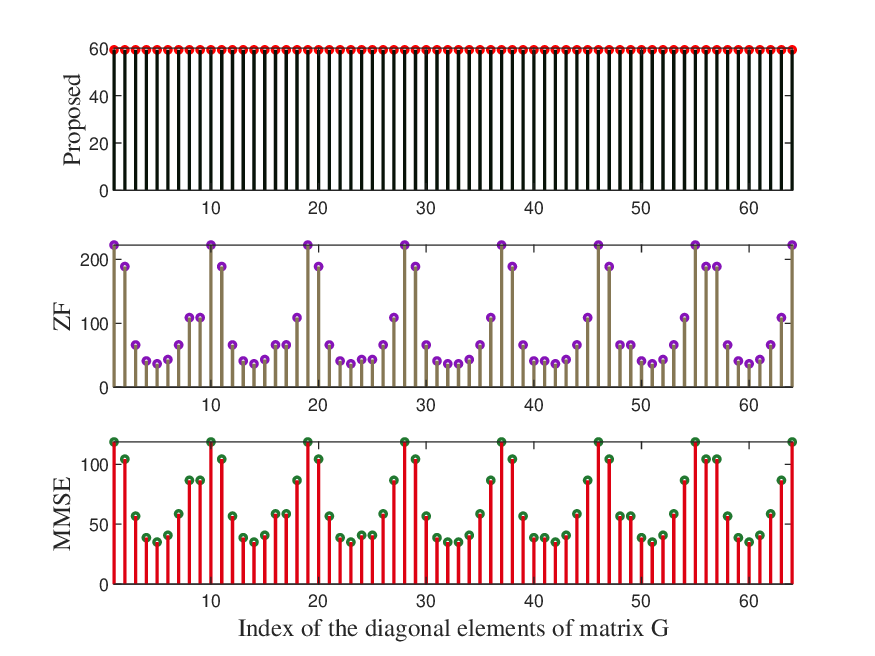}
    \caption{The diagonal elements of matrix $\mathbf{G}$.}
    \label{fig:enterDIA} 
\end{figure}

To further demonstrate the effectiveness of our proposed method,  we evaluate the SINRs of the received symbols, which are given by $ \mathrm{SINR}_i=\frac{1}{\left[\sigma_c^2\left(\zeta \sigma_c^2\mathbf{I}_{MN}+\mathbf{W}_\mathrm{DD}^{ {H}} \mathbf{H}_{\mathrm{c,DD}}^{{H}} \mathbf{H}_{\mathrm{c,DD}} \mathbf{W}_\mathrm{DD}\right)^{-1}\right]_{i,i}}-\zeta, 1\le i\le MN$, where $\zeta$ is an indicating factor, i.e., $\zeta=1$ for MMSE scheme and $\zeta=0$ for ZF scheme \cite{palomar2005minimum}. For ease of exposition, we denote the matrix $\mathbf{G}=\left(\zeta \sigma_c^2\mathbf{I}_{MN}+\mathbf{W}_\mathrm{DD}^{ {H}} \mathbf{H}_{\mathrm{c,DD}}^{{H}} \mathbf{H}_{\mathrm{c,DD}} \mathbf{W}_\mathrm{DD}\right)^{-1}$, whose diagonal elements are illustrated in Fig. \ref{fig:enterDIA} under the case of $P_T=30$ dBm. It is observed that the ZF scheme, compared to the MMSE scheme, has higher diagonal elements for the same index. Hence, the ZF scheme has a higher  $\mathrm{tr} (\mathbf{G})$, i.e., a poor average received symbol SINR, which causes the BER performance to decrease. On the other hand, our proposed design can equalize the SINRs of all the received symbols, such that all the diagonal elements of $\mathbf{G}$ are identical, as shown in the top of Fig. \ref{fig:enterDIA}, and consequently, the lower bound $P_e^{\mathrm{lb}}$ is achieved, as analyzed in (\ref{lb}).
\begin{figure}[t]
    \centering
\includegraphics[width=0.6\columnwidth]{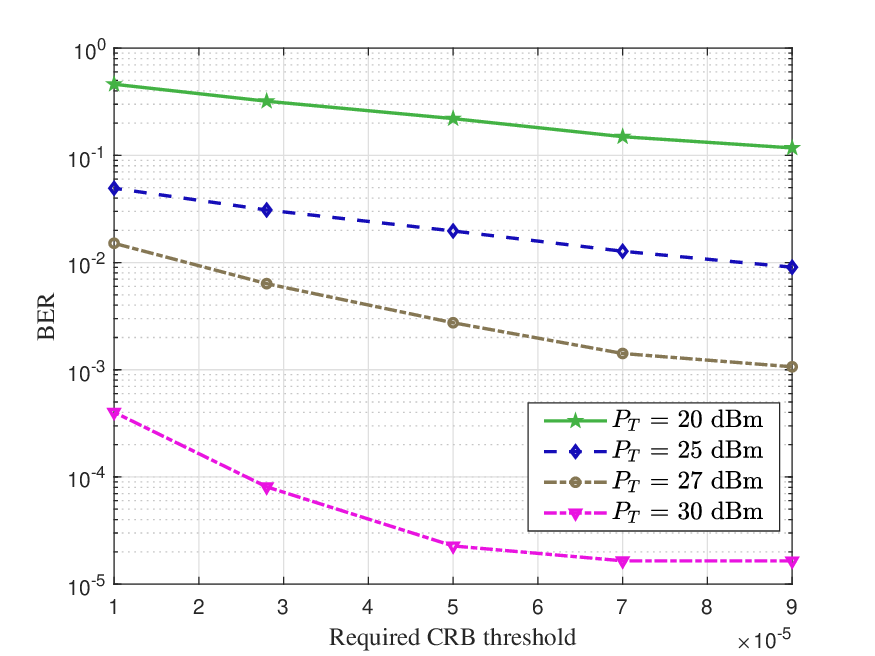}
\captionsetup{font=small,name=Fig, labelsep=period,justification=raggedright,singlelinecheck=false}
    \caption{The BER performance with various required BER thresholds.}
    \label{figcrb} 
\end{figure}
\begin{figure*}[t]
  \centering
 \begin{subfigure}{0.33\columnwidth}
  \centering
\includegraphics[width=\linewidth]{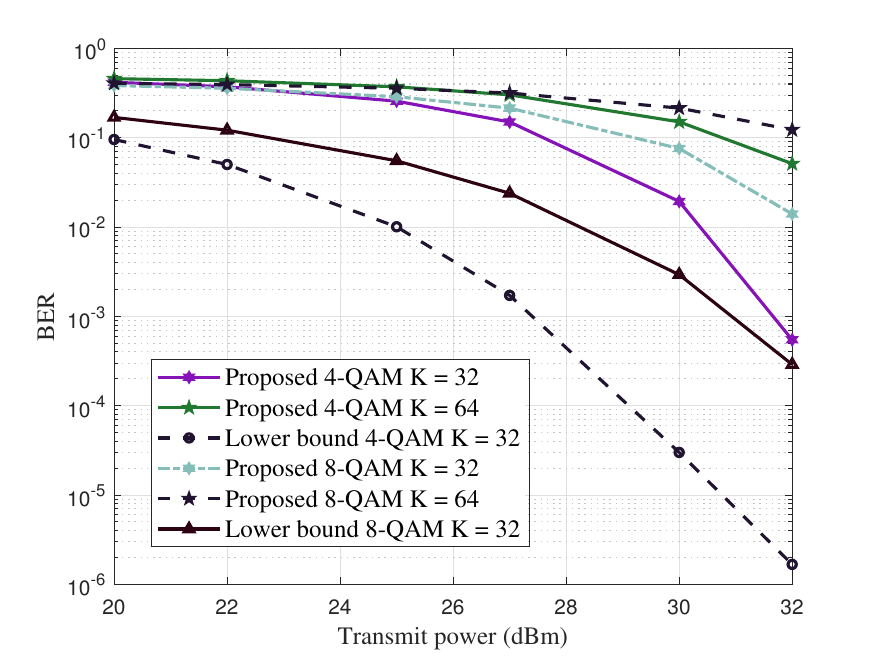}
       \caption{BER.}
    \label{figsubber}
  \end{subfigure}%
\begin{subfigure}{0.33\columnwidth}
  \centering
\includegraphics[width=\linewidth]{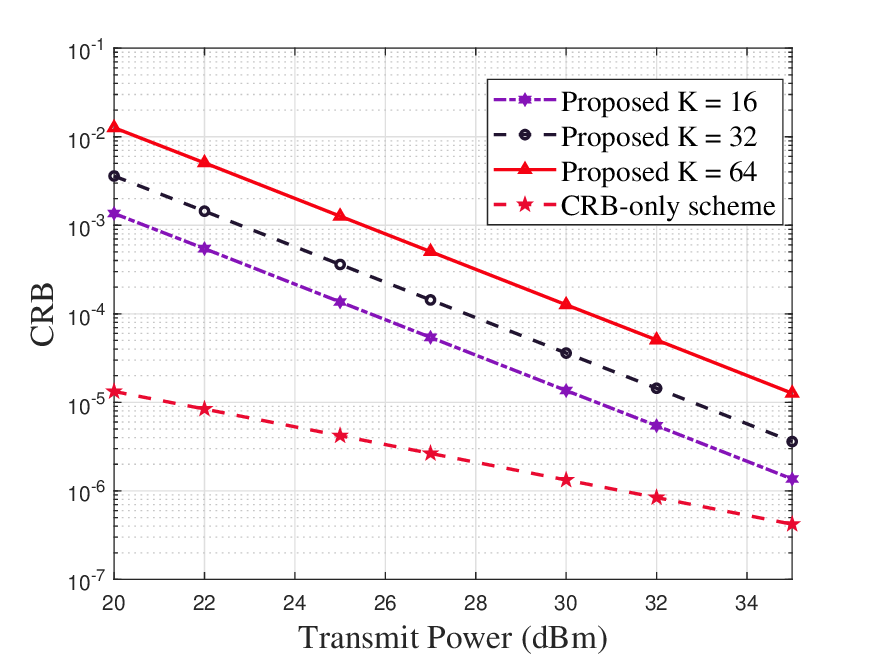}
    \caption{CRB.}
    \label{figsubcrb}
  \end{subfigure}
\begin{subfigure}{0.33\columnwidth}
    \centering
\includegraphics[width=\linewidth]{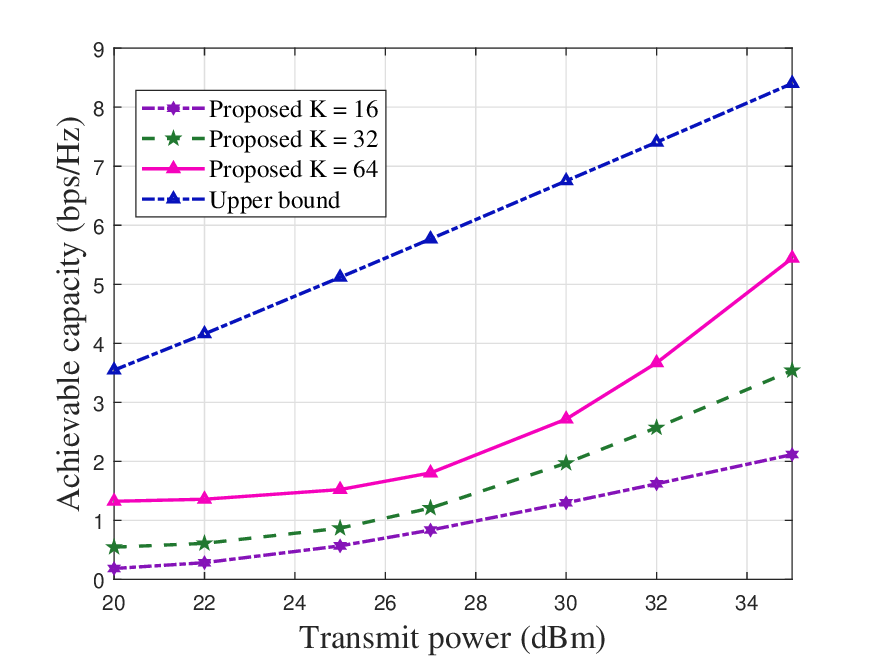}
    \caption{ Achievable capacity. }
    \label{figsubrate}
  \end{subfigure}
\captionsetup{font=small,name=Fig,labelsep=period,justification=justified,singlelinecheck=false}
  \caption{The system performance versus the transmit powers under various numbers of transmitted symbols. 
  % (a) The BER performance with different numbers of transmitted symbols; (b) The CRB performance with different numbers of transmitted symbols; (c) The achievable channel capacity of  
 }
   \label{figsys}
  \end{figure*}
% \begin{figure}
%     \centering
%     \includegraphics[width=\columnwidth]{Figures/BERnumber.eps}
%     \caption{The BER performance versus the number of transmitted symbols with $P_T=20$ dBm.}
%     \label{fig3}
% \end{figure}
% \begin{figure}
%     \centering
%     \includegraphics[width=\columnwidth]{Figures/CRB_num.eps}
%     \caption{The BER performance versus the number of transmitted symbols with $P_T=20$ dBm.}
%     \label{figCN}
% \end{figure}
% \begin{figure}
%     \centering
%     \includegraphics[width=\columnwidth]{Figures/Capacity_num.eps}
%     \caption{The BER performance versus the number of transmitted symbols with $P_T=20$ dBm.}
%     \label{figCAN}
% \end{figure}
 
Next, we explore the interplay between communication and sensing. We present the BER performance across varying the required CRB thresholds, as depicted in Fig. \ref{figcrb}. It is observed that as the required threshold decreases, the CRB constraint becomes more stringent, leading to a degradation in BER performance. Conversely, when the CRB constraint is loose, there exists a further improvement in the achievable BER performance as the system gains greater flexibility in power allocation for BER optimization. This phenomenon arises from the misalignment between the sensing and communication subspaces, which makes the desired precoder, for minimizing the BER, differ a lot from that satisfying the CRB thresholds. Moreover, in the case of $P_T=30$ dBm, it is interesting to see that the BER performance stays almost unchanged if the CRB thresholds were to be further increased. This is because the CRB constraint would become inactive at this point.

Recall from Corollary 1 that the BER and CRB are both optimal if there is only one data symbol to be transmitted in an OTFS block. In Fig. \ref{figsys}, we then investigate the system performance with different transmit powers in the context of various numbers of transmitted data symbols, denoted by $K, 1\le K\le MN$. For ease of study, we set the resolvable path $P=1$ for communication purposes at this time.  Fig. \ref{figsubber} illustrates the influence of the number of transmitted symbols on the BER performance, where the curves of lower bound are derived in (\ref{pem1}) in Appendix \ref{APPA} .  We can observe that all the considered schemes are likely to suffer BER performance degradation when the number of transmitted symbols increases.  This is expected since fewer transmitted symbols indicate better signal power in the received samples, which, in turn, enhances the BER performance. Besides, it is noted that the 4-QAM scheme demonstrates a significant BER outperformance compared to 8-QAM.  This discrepancy is caused by the fact that 4-QAM has a lower constellation density and possesses larger Euclidean distances between neighboring constellations, which enhances the precision of establishing distinct symbol decision regions. In Fig. \ref{figsubcrb}, we show the CRB performance with different transmitted symbol numbers, where the curve of 'CRB-only scheme' is obtained by solving problem (\ref{copr}). The optimal precoder for CRB-only criteria is a rank-1 matrix, which only transmits one symbol, having the best CRB performance. For our proposed design, the CRB performance decreases with the increase of $K$ because the precoder needs to allocate power to equalize the SINRs of all the received symbols, such that the BER performance is guaranteed. Moreover, it is observed that the proposed designs become closer to the CRB-only criteria in high SNR regimes as the CRB constraint is loosened.  
From the above discussions, we may conclude that the BER and CRB performance are enhanced in the context of utilizing fewer data symbols in one block, consequently improving transmission reliability and tracking accuracy.  However, it is noteworthy that this improvement comes at the cost of reduced spectral efficiency, as illustrated in Fig. \ref{figsubrate}.  The achievable capacity is calculated by $R=\frac{1}{MN}\log_2 \left|\mathbf{I}_{MN}+ \frac{ \mathbf{H}_{\mathrm{c,DD}} \mathbf{W}_\mathrm{DD}\mathbf{W}_\mathrm{DD}^{{H}} \mathbf{H}_{\mathrm{c,DD}}^{{H}}}{\sigma_c^2}\right|$ \cite{mohammed2021time} and the upper bound is obtained by maximizing $R$ given the transmit power. It is evident that transmitting more symbols can carry a more substantial amount of information, leading to achieving higher spectral efficiency. This is distinct from the BER and CRB designs and reveals a trade-off where transmission efficiency is sacrificed for increasing transmission reliability and sensing accuracy.
% To further characterize this trade-off, we show in Fig. \ref{figsubrate} the achievable capacity, denoted by $C=\frac{\eta}{MN}\log_2 \left| \mathbf{I}_{MN}  + \frac{{\mathbf{H}}_\mathrm{c,DD} \mathbf{W}_\mathrm{DD}\mathbf{W}_\mathrm{DD}^{H}{\mathbf{H}}_\mathrm{c,DD}^{{H}} }{\sigma_c^2} \right|$ with $\eta$ being $K/MN$. 

 \section{Conclusions}
\label{sec6} 
 In this paper, we devised the minimum BER precoder for OTFS-ISAC systems, accounting for the constraints imposed by the maximum available transmission power and the required sensing performance. We first formulated the desired problem, which optimizes the lower bound of the average BER. To circumvent the non-convexity of the formulated problem, we proposed to adopt the SVD approach to separate the original problem into two independent sub-problems. Then, the feasible conditions have been analyzed, followed by the determination of the low-complexity precoder by focusing on the Lagrangian dual problem utilizing the ellipsoid method. Numerical results demonstrated that our proposed precoder can provide considerable BER performance gain compared to traditional ZF and MMSE equalizers and evaluated the non-trivial trade-off between sensing and communications for the dual-functional precoder design.

\begin{appendices}
\section{Proof of Corollary 1} \label{APPA}
\vspace{-0.2cm}
Based on Theorem 1, we know that the optimal precoder $\mathbf{W}_\mathrm{s}^*$ for CRB-only criteria only transmits one single symbol. We then prove that the BER performance is also optimal in the case of transmitting one single symbol. For ease of exposition, we first denote $\mathbf{\Lambda}_\mathrm{c} =\mathrm{diag}(\Lambda_1, \cdots, \Lambda_{MN})$ with descending order. Hence, the diagonal elements of $\mathbf{\Lambda}_\mathrm{c}^{-1}$ are given by $\Lambda_i^{-1}, i=1,2\cdots, MN$, where $\Lambda^{-1}_1\le\Lambda_2^{-1}\le \cdots \Lambda_{MN}^{-1}$.  Relying on (\ref{pem}), the BER lower bound for BER-only criteria transmitting $K (1\le K\le MN)$ symbols can be straightforwardly derived by
\begin{align}
    P_{e,\min}^\mathrm{lb}(K)=\alpha \mathcal{Q}\left(\sqrt{\frac{\beta K P_T}{\sigma_c^2\left(\operatorname{tr}\left( \mathbf{\Lambda}_{\mathrm{c},K}^{-1/2}\right)\right)^2}} \right) \label{pem1},
\end{align}
where $\mathbf{\Lambda}_{\mathrm{c},K}^{-1/2}=\operatorname{diag}(\Lambda_1^{-1/2},\cdots,\Lambda_{K}^{-1/2})$ includes the smallest $K$ square roots of diagonal elements in $\mathbf{\Lambda}_\mathrm{c}^{-1}$. 
Based on the mathematical induction approach, it is readily shown that (\ref{pem1}) achieves the minimum value when $K=1$. Owing to the directness of the proof and page limitation, the detailed derivations are omitted here. Next, we consider the optimal dual functional precoder when $K=1$. The single OTFS-ISAC symbol should be transmitted through the strongest eigen sub-channel. In this case, the precoder is reduced to a vector $\mathbf{w}_\mathrm{dd}$, such that problem 
(\ref{svdpro}) can be rewritten as 
% \begin{subequations}
% \begin{align}
% \max_{\mathbf{w}_\mathrm{dd}} \quad &\mathbf{w}_\mathrm{dd}^H {\mathbf{H}}_\mathrm{c,DD}^H{\mathbf{H}}_\mathrm{c,DD}\mathbf{w}_\mathrm{dd} \\
% \mathrm{s.t.} \quad &\mathbf{w}_\mathrm{dd}^H {\mathbf{\dot{H}}}_\mathrm{s,DD}^H{\mathbf{\dot{H}}}_\mathrm{s,DD}\mathbf{w}_\mathrm{dd} \ge \gamma_1, \quad || \mathbf{w}_\mathrm{dd}||\le P_T.
% \end{align}
% \end{subequations}
%  Thus, the above problem is equivalent to
\begin{subequations} \label{fimpro}
\begin{align}
\max_{\mathbf{w}_\mathrm{dd}} \quad & \Lambda_1| \mathbf{w}_{\mathrm{c},1}^H\mathbf{w}_\mathrm{dd}|^2  \\
\mathrm{s.t.} \quad & \Xi_1| \mathbf{w}_{\mathrm{s},1}^H\mathbf{w}_\mathrm{dd}|^2 \ge \gamma_1, \quad || \mathbf{w}_\mathrm{dd}||^2\le P_T,
\end{align}
\end{subequations}
where $\mathbf{w}_{\mathrm{c},1}$ is the eigenvector corresponding to $\Lambda_1$. In the case of the CRB constraint being inactive, it can be obtained that $\mathbf{w}_\mathrm{dd}=\sqrt{P_T}\mathbf{w}_{\mathrm{c},1}$, i.e., allocating entire power to align $\mathbf{w}_{\mathrm{c},1}$. When the CRB constraint is active, it becomes evident that $\mathbf{w}_\mathrm{dd} \in \mathrm{span}\left\{\mathbf{w}_{\mathrm{c},1}, \mathbf{w}_{\mathrm{s},1}  \right\}$ \cite{liu2021cramer}. Note that $||\mathbf{w}_{\mathrm{c},1}||=||\mathbf{w}_{\mathrm{s},1}|| =1$. Consequently, $\mathbf{w}_\mathrm{dd}$ can be represented by $\mathbf{w}_\mathrm{dd}=x\mathbf{w}_{\mathrm{s},1}+y \mathbf{w}_{\mathrm{u}}, x,y \in \mathbb{C}$, 
% \begin{align}
%\mathbf{w}_\mathrm{dd}=x\mathbf{w}_{\mathrm{s},1}+y \mathbf{w}_{\mathrm{u}},  \quad x,y \in \mathbb{C},
% \end{align}
where $\mathbf{w}_{\mathrm{u}} =\frac{\mathbf{w}_{\mathrm{c},1}-(\mathbf{w}_{\mathrm{s},1}^H\mathbf{w}_{\mathrm{c},1})\mathbf{w}_{\mathrm{s},1}}{||\mathbf{w}_{\mathrm{c},1}-(\mathbf{w}_{\mathrm{s},1}^H\mathbf{w}_{\mathrm{c},1})\mathbf{w}_{\mathrm{s},1}||}$ is orthogonal to $\mathbf{w}_{\mathrm{s},1}$. Then, problem (\ref{fimpro}) can be rewritten as 
\begin{subequations}
\begin{align}
\max_{x,y} & \quad \left|x \mathbf{w}_{\mathrm{c},1}^H \mathbf{w}_{\mathrm{s},1}+y \mathbf{w}_{\mathrm{c},1}^H \mathbf{w}_\mathrm{u}\right|^2 \\
\mathrm { s.t. } &  \quad \left|x\right|^2|\Xi_1|=\gamma_1, \ \left|x\right|^2+\left|y\right|^2=P_T.
\end{align}
\end{subequations} 
It is evident that $|x|^2=\frac{\gamma_1}{|\Xi_1|}$ and $|y|^2=P_T-\frac{\gamma_1}{|\Xi_1|}$. Thus, we have $x=\sqrt{\frac{\gamma_1}{|\Xi_1|}} \frac{\mathbf{w}_{\mathrm{s},1}^H\mathbf{w}_{\mathrm{c},1}}{|\mathbf{w}_{\mathrm{s},1}^H\mathbf{w}_{\mathrm{c},1}|}$ and $y=\sqrt{P_T-\frac{\gamma_1}{|\Xi_1|}} \frac{\mathbf{w}_{\mathrm{u}}^H\mathbf{w}_{\mathrm{c},1}}{|\mathbf{w}_{\mathrm{u}}^H\mathbf{w}_{\mathrm{c},1}|}$ to align $\mathbf{w}_{\mathrm{c},1}^H \mathbf{w}_{\mathrm{s},1}$ and $ \mathbf{w}_{\mathrm{c},1}^H \mathbf{w}_\mathrm{u}$, respectively, such that the objective value achieves the maximum value, which completes the proof. 
\end{appendices}
\bibliographystyle{IEEEtran}
\bibliography{citation}
\end{document}